\documentclass[a4paper,fleqn,usenatbib]{mnras}

\usepackage{newtxtext,newtxmath}
\usepackage[T1]{fontenc}
\usepackage{ae,aecompl}

\usepackage{amsmath}	% Advanced maths commands
\usepackage{amssymb}	% Extra maths symbols
\usepackage{bm}
\usepackage{here}
\usepackage[dvipdfmx]{graphicx}
\usepackage{bmpsize}
\usepackage{color}

\title[The thermal-radiative wind in H~1743-322]{The thermal-radiative wind in low mass X-ray binary H~1743-322; Radiation hydrodynamic simulations}

\author[Tomaru et al.]{
\thanks{E-mail: ryota.tomaru@ipmu.jp}
Ryota Tomaru,$^{1,2}$
Chris Done,$^{3,2}$
Ken Ohsuga,$^{4}$
Mariko Nomura$^{5}$ and
\newauthor
Tadayuki Takahashi$^{2,1}$ 
\\
$^{1}$Department of Physics, The University of Tokyo, 7-3-1 Hongo, Bunkyo, Tokyo 113-0033, Japan \\
$^{2}$Kavli Institute for the Physics and Mathematics of the Universe (WPI), University of Tokyo, Kashiwa 277-8583, Japan\\
$^{3}$Centre for Extragalactic Astronomy, Department of Physics, University of Durham, South Road, Durham DH1 3LE, UK\\
$^{4}$Center for Computational Sciences, University of Tsukuba, 1-1-1 Ten-nodai, Tsukuba, Ibaraki 305-8577,
Japan\\
$^{5}$Faculty of Natural Sciences, National Institute of Technology, Kure College, 2-2-11 Agaminami, Kure, Hiroshima 737-8506, Japan
}

\date{Accepted XXX. Received YYY; in original form ZZZ}

\pubyear{2015}

\begin{document}
\label{firstpage}
\pagerange{\pageref{firstpage}--\pageref{lastpage}}
\maketitle

\begin{abstract}
Blueshifted absorption lines are seen in  
high inclination black hole binary systems 
in their disc dominated states, 
showing these power an equatorial disc wind. 
While some contribution from magnetic winds remain a possibility,
thermal and thermal-radiative winds are expected to be present.
We show results from radiation hydrodynamic simulations 
which show that the additional radiation force 
from atomic features (bound-free and lines) are important along with electron scattering.
Together, these increase the wind velocity at high inclinations, 
so that they quantitatively match the observations in H~1743-322,
unlike purely thermal winds which are too slow.
We highlight the role played by shadowing of the outer disc from the (sub grid) inner disc Compton heated layer,
and show that the increase in shadow from the higher 
Compton temperature after the spectral transition 
to the hard state  leads to strong suppression of the wind.
Thermal-radiative winds explain all of the spectral features 
(and their disappearance) in this simplest wind system and magnetic winds play only a minor role.
We speculate that thermal-radiative winds 
can explain all the spectral features seen 
in the more complex (larger disc size) binaries, 
GRO~J1655-40 and GRS~1915+105, without requiring magnetic winds. 
\end{abstract}


%%%%%%%%%%%%%%%%%%%%%%%%%%%%%%%%%%%%%%%%%%%%%%%%%%

%%%%%%%%%%%%%%%%% BODY OF PAPER %%%%%%%%%%%%%%%%
%%

\section{Introduction}
Outflowing gas is ubiquitous in accreting systems across mass scale
from protostars (e.g.\citealt{Bachiller1996,Carrasco-Gonzalez2010}) through to both stellar
mass compact objects in X-ray binaries (e.g. \citealt{Remillard2006,DiazTrigo2016}) 
and supermassive black holes in active galactic nuclei (AGN) 
at the centre of galaxies (e.g. \citealt{Heckman1990,Hagino2016}).  
These outflows, as either highly collimated jets (e.g. \citealt{Burrows1996,Mirabel1999,Remillard2006})
or wider angle winds (e.g. \citealt{Tombesi2010b,Ueda2009}), 
can have a large mass and/or kinetic power so they can have a significant
impact on the accretion environment. The important open questions in 
the field are how these outflows are driven, whether there are common 
processes across these very different objects, 
and whether there is a causal relation between jets and disc winds. 
Here we address these questions in the specific setting of the
Galactic Compact objects.

Low-mass X-ray binary systems, both neutron star binaries (NSBs) and
black holes binaries (BHBs) can show blueshifted absorption lines
from highly ionised ions. These winds are fairly slow, typically less
than 1000 km/s \citep{Miller2015}, seen preferentially in high
inclination systems (e.g. the compilation by \citealt{Ponti2012}) where
the binary separation is large (e.g. \citealt{DiazTrigo2016}).These
features indicate an equatorial disc wind launched from large radii,
where the latter is indicated both by the wind preference for large
systems and its slow velocity. 
The final observational constraint is that the winds are seen only when
the spectra are soft and fairly luminous, in disc dominated states,
rather than in the Compton dominated hard state \citep{Ponti2012,Ponti2014}. 
 The spectral change can be explained as a transition
between a disc and hot accretion flow, where the collapse of the
larger scale height hot flow into a geometrically thin disc also
triggers the collapse of the steady compact radio jet (see e.g. the
review by \citealt{Done2007b}). This association of the
onset of the wind with the suppression of the jet was first noted in
GRS~1915+105 by \citet{Neilsen2009}, who suggested that 
the same magnetic field responsible for the jet in the hard state
underwent a reconfiguration to power a wind by Lorentz force (magnetic wind) in the soft state.
This is consistent with the general disappearance of the
wind in the hard states (see e.g. \citealt{Ponti2012}), though not with
the simultaneous observation of jets and winds in more complex
intermediate states (e.g. \citealt{Kalemci2016,Homan2016}), nor
does it explain the preference of winds for systems with larger discs.
Also, the current magnetic wind models \citep{Fukumura2010} require a special configuration of the large scale magnetic fields 
threading accretion disc, but the existence of such magnetic fields is unknown.

Instead, thermal winds have a clear link to both disc size and
illuminating spectrum and luminosity. The pioneering work by
\citet[hereafter B83]{Begelman1983a} showed that bright X-ray emission from the
inner disc and corona heats the surface of the disc at larger radii
to the Compton temperature, $T_\mathrm{IC}$. %=$$ \int h\nu F_{\nu}d\nu/(4k \int F_{\nu}d\nu)$.  
This temperature depends only on the spectrum of the radiation, so the 
entire disc surface has a heated atmosphere at this temperature, 
which is unbound at radii $R_\mathrm{IC}$
where this temperature means that particles have enough energy to
escape i.e.$ kT_\mathrm{IC}=GM_c\mu m_p/R_\mathrm{IC}$ 
where where $\mu$ is the mean molecular weight, set to $0.61$.
This launches a thermal wind at radii larger than the
inverse Compton radius, $R_\mathrm{IC} \sim (6.4\times 10^4 / T_\mathrm{IC,8} )~R_g$
($R_g=GM_c/c^2,~T_\mathrm{IC,8}= T_\mathrm{IC}/10^8~\mathrm{K}$), giving a clear
dependence on both the spectrum (through $T_\mathrm{IC})$ and the disc size,
as observed.

However, such thermal wind solutions were disfavoured due to a single,
highly anomalous, wind from GRO~J1655-40, where the wind launch radius
could be inferred from density diagnostics. The observed ionisation
state was unusually low for these winds, at
$\xi=L/nR^2\sim 100$, while the density measured from metastable
levels was quite high.  The observed luminosity was fairly low, so the
derived launch radius was much smaller than $R_\mathrm{IC}$, ruling out a
thermal wind \citep{Miller2006,Kallman2009,Luketic2010,Neilsen2012b,Higginbottom2015}. 
However, the unusual properties of the broad-band continuum seen during this
observation indicate that the intrinsic source luminosity may be
substantially underestimated, with the wind being so strong that it
has become optically thick \citep{Done2007b,Uttley2015,Neilsen2016,Shidatsu2016}.

\citet[hereafter D18]{Done2018} revisited the thermal
wind models of B83, combining their analytic equations for the mass
loss rate with the hydrodynamic results of \citet{Woods1996} to
predict the wind observables of column density, ionisation state
and velocity as a function of spectral shape, luminosity and inclination. They
stressed that the observed winds are generally seen at fairly high
$L/L_\mathrm{Edd}$, and suggested that radiation force could play an
important role, so that the winds are more accurately described as
thermal-radiative rather than simply thermal.

Here we show the first hydrodynamic simulations of thermal-radiative winds with realistic illuminating spectra and full opacities. 
We tailor these simulations for the typical BHB system H~1743-322,
where Chandra grating data clearly detect the wind during a soft state, 
but show only tight upper limits on similar absorption features during a hard
state.  We show that the wind velocity in the soft state is
much faster than predicted by purely thermal winds (see also~
\citealt{Luketic2010,Higginbottom2015}), but that the radiation force
from the combination of electron scattering and atomic opacities (both
edges and lines) produces additional driving which increases the wind
velocity at high inclinations, matching well to that observed.  We show that
the strong suppression of the wind when the system switches to the
hard state is predicted by the thermal-radiative wind
models as claimed in D18, but for a rather different reason, 
more connected to radiation
transfer effects in the inner disc atmosphere than to the different
outer disc wind properties. These simulations demonstrate that thermal-radiative
winds are the origin of most (perhaps all) of the outflows seen
in Galactic compact binaries.

\section{The BHB H~1743-322}

There are currently four BHBs which show wind features in their soft
states, namely GRO~J1655-40, GRS~1915+105, H~1743-322 and 4U~1630-472
(see e.g. the compilation of \citealt{Ponti2012}).  The anomalous
features of the wind in one observation of GRO~J1655-40 are discussed
above (though see \citealt{Higginbottom2018}) for a thermal wind simulation of a more normal soft state.
GRS~1915+105 has a mean luminosity around Eddington, so radiation force will be extremely important.
It also typically shows complex spectra rather than standard soft or hard states, 
and has a truly enormous disc so needs a very large simulation domain. 
The system parameters for 4U~1630-472 are not well
known and it has a very large galactic column which makes determining the soft state 
disc luminosity difficult. 

This leaves H~1743-322 as the best object to model in detail.
This object has recurrent transient outbursts, switching from
hard to soft on the fast rise, and soft to hard on the slower
decline. Chandra High Energy Transmission Grating (HETG) observations
during a standard disc dominated state in 2003 show blueshifted absorption lines from Fe {\scriptsize XXV} and Fe {\scriptsize XXVI}
(ObsID:3803, see \citealt{Miller2006b}), with only upper limits on these
absorption lines in the hard state in 2010 (ObsID:11048, see \citealt{Miller2012}).
We reanalyse the data of soft state, and fit with Gaussian absorption line %show the region near the iron
({\sc gabs} in {\sc xspec}) to get columns and velocity.
We find these values are within $\pm 1 \sigma$ errors of those of previously 
reported i.e the column and velocity of Fe {\scriptsize XXV},
$N_\mathrm{XXV}=8.2^{+1.2}_{-1.1}\times 10^{16}$~cm$^{-2}$ and $V_\mathrm{XXV} = 370 \pm 120 ~\mathrm{cm^{-2}}$,  while those of Fe {\scriptsize XXVI} are
$N_\mathrm{XXVI}=4.0\pm 0.3\times 10^{17}$~cm$^{-2}$ and $V_\mathrm{XXVI} = 630^{+80}_{-120} ~\mathrm{km/s}$.

We adopt system values from \citet{Shidatsu2019}, where
$M=7M_\odot$, $D=8.5$~kpc, and inclination of $75\pm3^\circ$ \citep{Steiner2012}.
We note that these are different to those assumed in D18, so that the observed bolometric  
luminosities from the RXTE PCA data for the soft and hard state
correspond to $L/L_\mathrm{Edd}=0.3$, and $0.06$ (compare to $0.1$ and $0.02$ in D18).

The remaining key parameter which determines the strength of a thermal wind is the disc size, $R_\text{disc}$. 
This is quite poorly known. 
D18 assume $R_\text{disc}=5 R_\text{IC}$ to connect to the hydrodynamic simulations of \citet{Woods1996} but \citet{Shidatsu2019} note that the outburst frequency puts the source on the edge of the disc instability, similarly to GX339-4.
We use the MAXI lightcurve to measure the mean X-ray luminosity over a 10-year time span.
The mean MAXI 2-20 keV count rate is about 10 times lower than 
the count rate seen during the low/hard state ($L/L_\text{Edd} = 0.06$) used by \citet{Shidatsu2019}, 
%which is at $L/L_\text{Edd} = 0.06$ for the assumed black hole mass and distance. 
This converts to a mean mass transfer rate of $\sim 6\times 10^{16}\mathrm{~g~s^{-1}}$, where the disk instability condition predicts an orbital period of $\sim 7$ hours \citep{Coriat2012}. 
Assuming the companion star has a similar mass ratio as in GX339-4 then this implies a Roche lobe size of $1.7\times 10^{11}$~cm, and hence a disc radius of $1.2\times 10^{11}$~cm ($70\%$ Roche lobe size), which is $0.2~R_\text{IC}$. 
This is the point at which thermal winds are launched, so the wind will depend quite sensitively on size scale here. Given the uncertainties, we start first by calculating the wind produced by a disc with $R_\text{disc}=R_\text{IC}$, and then explore the effect of changing parameters. 
\if0
\begin{figure}
\includegraphics[width=0.9\hsize]{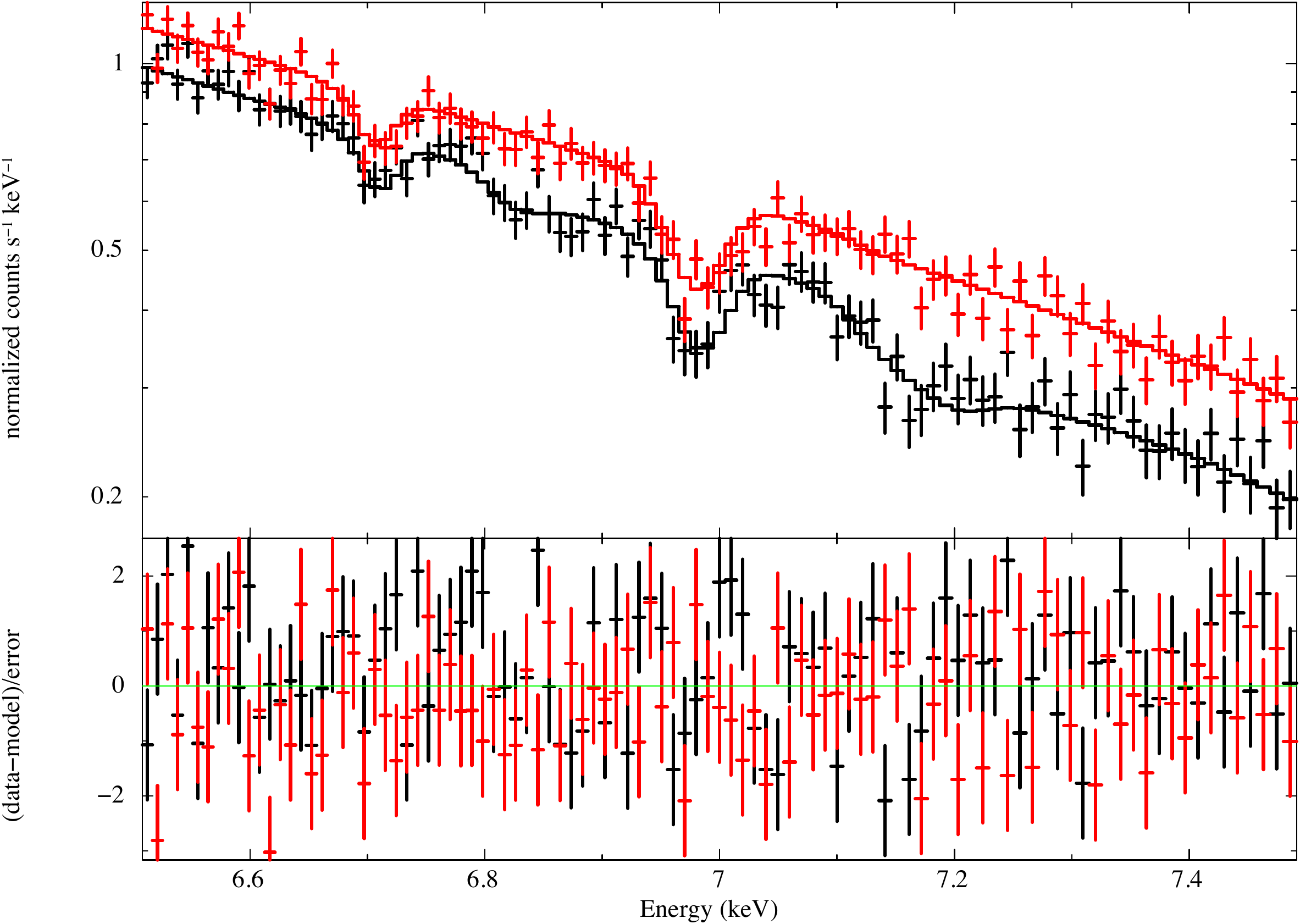}
\caption{The spectrum of H~1743-322 fitting with {\sc diskbb+pow+gabs+gabs+gabs+gabs} model. }
\label{fig:obs}
\end{figure}
\fi

 \section{Computational Method}

\subsection{Code Overview}
 
 \begin{figure*}
  \includegraphics[width=\hsize]{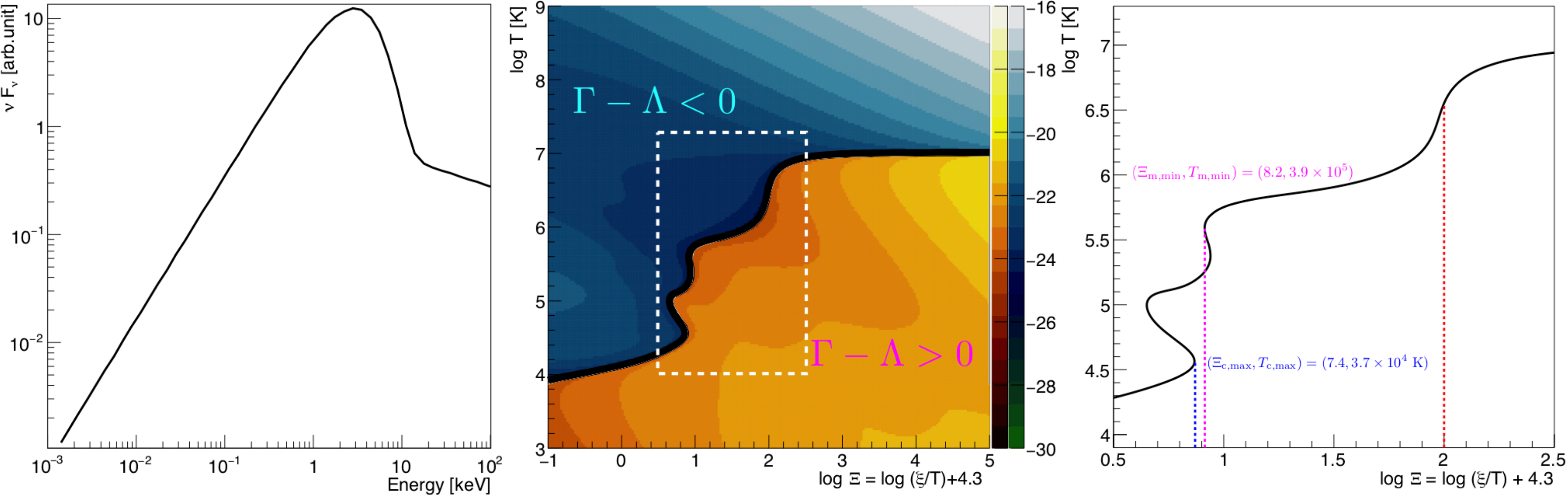}
  \caption{a) The energy flux of H~1743-322  which consists of a disc-blackbody plus Comptonised power-law ({\sc diskbb+diskbb*simple} in {\sc xspec}).
  b) The heating and cooling function ($\Gamma - \Lambda$ ). The black curve shows the thermal equilibrium ($\Gamma - \Lambda$ = 0), the right colour bar shows the absolute  logarithm value of heating minus cooling function ($ \mathrm{log} |\Gamma - \Lambda|$). The Compton temperature of this spectrum is $T_\mathrm{\mathrm{IC}}=1.0\times 10^7~\mathrm{K}$. 
  c)The zooming figure of b) surrounding by white dashed line. The vertical lines show the instability points. Red vertica lines is $\Xi_\mathrm{h,min}=100$.   }
  \label{fig:continuum}
 \end{figure*}
 
 \begin{figure}
  \includegraphics[width=0.9\hsize]{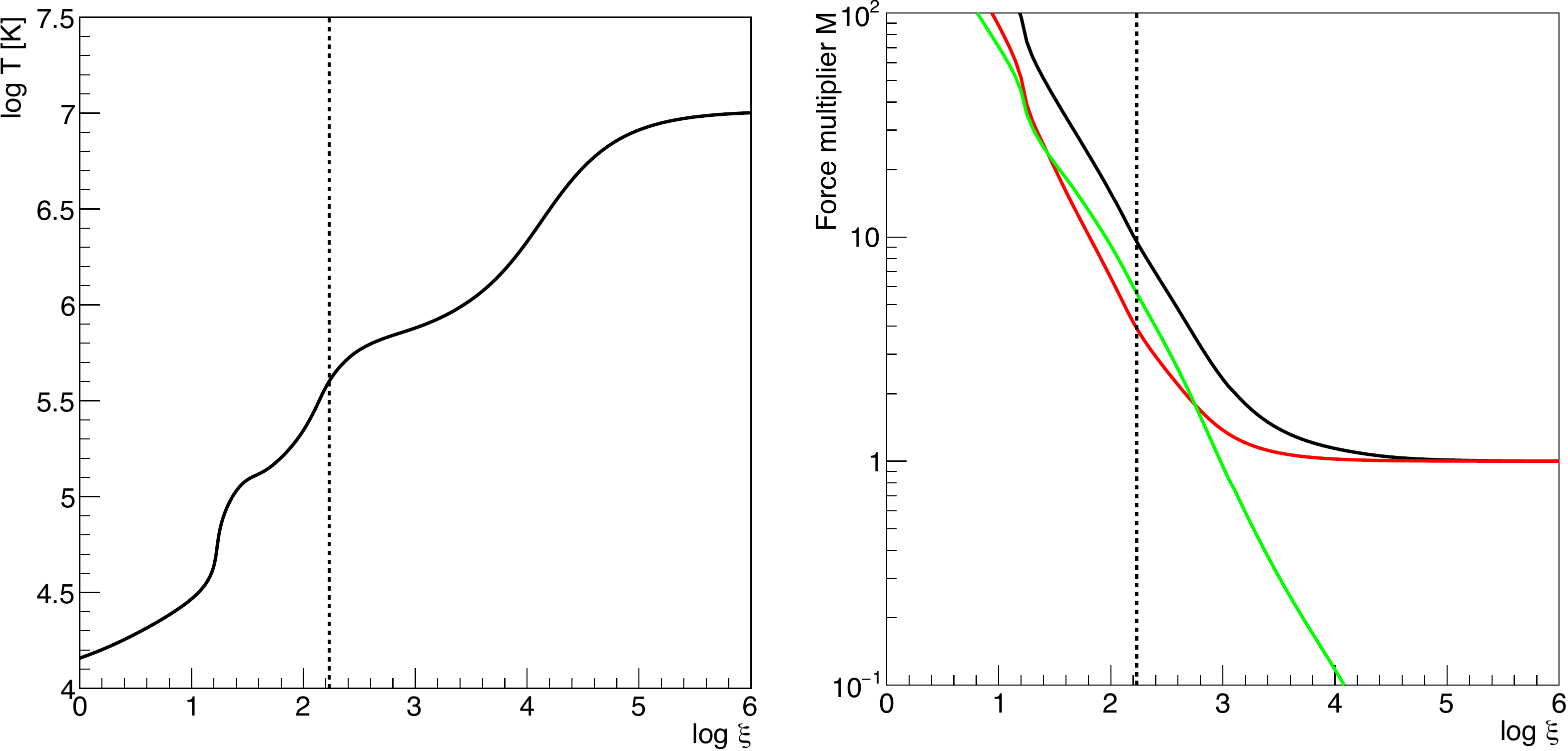} 
  \caption{a)Thermal equilibrium curve of $\log \xi \mathrm{vs} \log T $.
   b): Force multiplier of continuum process(bound-free plus scattering, red), line (green) and total (black) at the thermal equilibrium state.
  The vertical dashed line shows $\xi_\mathrm{H,min} = 1.7\times10^2$ which is the ionisation parameter of disc surface.
  }
  \label{fig:force m}
 \end{figure}

%We use the 
Our radiation-hydrodynamic simulation code was developed by \citet{Takahashi2013}, 
described most recently in \citet{Nomura2016}.
This is similar to the code used in earlier work by \citet{Proga2000,Proga2004}. 
We solve the standard equations of motion in spherical polar coordinates, including the radiation force
on the material, together with the continuity equations for mass and energy equation, where the latter includes a
net heating/cooling term for the material, $\mathcal{L}$. This is defined by 
 \begin{equation}
 \rho \mathcal{L} = \left( \frac{\rho}{\mu m_p} \right)^2 [\Gamma(\xi, T) -\Lambda(\xi, T)]
 \end{equation}
where mass density and number density are related by$\rho=n \mu
m_p$. The heating, $\Gamma$, and cooling, $\Lambda$, rates per unit
volume are calculated by {\sc cloudy} \citep{Ferland2003}, although the 
simple heating and cooling functions are employed in \citet{Nomura2016}.
These depend on the shape of the illuminating spectrum, which is discussed
in more detail in the next section, as well as on the
temperature $T$ and ionisation parameter
\begin{equation}
\xi(R,\theta) = \frac{L_x}{n R^2} = \frac{4\pi \mu m_p F_\mathrm{x}}{\rho}
\end{equation}
where $F_x$ is the X-ray flux and $L_x$ is the X-ray luminosity at each grid point. These 
are defined from the intrinsic luminosity, $L_\mathrm{x0}$, which is 
modified for absorption/scattering along the line of sight so that $L_x=L_\mathrm{x0}e^{-\tau(R,\theta)}$. 
We assume that the corona and inner disc are both effectively a central point source, so that 
\begin{equation}
 \tau (R, \theta) = \sum  M(\xi,T) \kappa_\mathrm{es} \rho(R, \theta)  \delta R +\tau_0(\theta)
 \label{eq:tau}
\end{equation}
where $\kappa_\mathrm{es}=0.34 ~\mathrm{cm^2 ~g^{-1}}$ is the mass-scattering coefficient for free electrons,
$\tau_0$ is optical depth of inner corona discribed in Sec.\ref{sec:inner_corona} and 
 $M(\xi, T)$ is the force multiplier defined by \citet{Tarter1973}.
This force is calculated by 
{\sc cloudy} and accounts for the additional opacity from continuum process (bound-free; photoelectric absorption edges) and line process (bound-bound transitions) (Fig.\ref{fig:force m}b),  but does not 
include the additional force from velocity shifts as the velocity is low.
This also defines the radiation force on the material as $f_\mathrm{\mathrm{rad}} = \frac{\kappa_\mathrm{es} M(\xi, T)}{c} F_x ~\mathrm{[cm~s^-2]}$.
{\sc cloudy} also calculates the ion fractions of Fe {\scriptsize XXV} and 
Fe {\scriptsize XXVI}, $f_\text{XXV} (\xi, T)$ and  $f_\text{XXVI}(\xi, T)$ at each point of $\xi$ and $T$.

We adopt an irradiated disc shape, where $H_\mathrm{d}/R \propto R^{2/7}$  
(\citealt{Cunningham1976}, as recast by \citealt{Kimura2019}) where
\begin{equation}
\begin{split}
    H_\mathrm{d}/R& = 1.5 \times 10^{-3} \left(\frac{L}{L_\mathrm{\mathrm{Edd}}}\right)^{1/7} \left(\frac{M_c}{M_\odot}\right)^{-1/7}\left(\frac{R_\mathrm{out}}{R_\mathrm{g}}\right)^{2/7}\left(\frac{R}{R_\mathrm{\mathrm{out}}}\right)^{2/7}\\
         & =f_d \left(\frac{R}{R_\mathrm{out}}\right)^{2/7}
\end{split}
\label{eq:scale_d}
\end{equation}
In the simulation grid, we set angle of disk from mid-plane $\alpha_d (R) = 
\arctan \{ H_d(R)/R-H_d(R_\text{in})/R_\text{in}\}$ in order to set $\alpha_d = 0~(\theta = \pi/2)$ 
at inner radius of computational domain $R_\text{in}$ 
(see also, Appendix.\ref{sec:code}). 

We apply the axially symmetric boundary at the rotational axis of the accretion disk, $\theta= 0$ so that $\rho, p$ and $v_r$ are symmetric, while $v_\mathrm{\theta}$ and $v_ {\phi}$ are antisymmetric). We apply a reflecting boundary at $\theta = \pi/2 $ so that $\rho, p, v_r$ and $v_\phi$ are symmetric, but $v_\mathrm{\theta}$ is antisymmetric. Outflow boundary conditions are employed at the inner and outer radial boundaries, so that 
matter can freely leave but not enter the computational domain. We set our radial computational grid from 
$R=(0.01-1)R_\mathrm{out}$, with 120 values logarithmically spaced. 
Our assumed value of $R_\text{out}=6.6\times 10^5R_\text{g}$ is coincidentally equal to $R_\text{IC}$ for the Compton temperature of the soft state SED of $1.0\times 10^7$~K.

We set a specific polar angle grid to follow the shape of the disc surface,
so we take 120 values of $\theta$ to cover the disc/atmosphere transition for angles 
$\theta > \pi/2-\alpha_d(R_\text{out})$
and another 120 values to cover the wind region with constant 
solid angle $d(\cos\theta)=\sin\theta d\theta$. 
In the disc zone and disk surface, we set $v_R =0, v_\phi=v_k = \sqrt{GM_c/R}\sin\theta$ at each time step. 

 We initially set temperature $T(R,\theta)=1.1 \times 10^7 (R/6R_\text{g})^{-3/4} \mathrm{K}$, 
density $\rho_0 (R,\theta) = 1.0\times 10^{-33} \mathrm{g~cm^{-3}}$ (except in the disk region), $v_R(R,\theta)=0$,
$v_\theta(R, \theta)=\sqrt{GM/R}\sin\theta$ and $v_\phi(R,\theta)=0$. 

For comparison with observation, we calculate the ion density of Fe {\scriptsize XXVI}  and Fe {\scriptsize XXV}
from simulations via $n_\text{XXVI}=n~A_\text{Fe} f_\text{XXVI}(\xi, T)$ 
and $n_\text{XXV}=n~A_\text{Fe} f_\text{XXVI}(\xi, T)$, where $A_\text{Fe}=3.3\times10^{-5}$ 
is iron abundance relative to hydrogen which is the same value as \citet{Miller2006b}.
We use these to define the ion column density along any line of sight, 
and to calculate the column density weighted mean velocity for each ion.

\subsection{Ionisation state calculations}

We focus first on the soft state, where the absorption lines are
detected, and show the SED from quasi-simultaneous RXTE PCA data
(ObsIDs P95368-01-01-00) in Fig. \ref{fig:continuum}a.  We use this to calculate the
heating and cooling rates from {\sc cloudy}, assuming that the illuminated
gas is optically thin. Fig.\ref{fig:continuum}b shows the heating and cooling rates for
material at a given temperature and pressure ionisation parameter
$\Xi=P_\mathrm{rad}/P_\mathrm{gas}= L/(4\pi R^2 c nkT) =(\xi/T) \times 1/(4\pi c k)$ where
$\xi$ is the standard ionisation parameter defined above.  
Thermal equilibrium is produced when the heating rate per unit volume is equal
to its cooling rate, as indicated by the heavy black line in Fig.\ref{fig:continuum}b. 
The blue region to the left of this curve shows that the cooling rates are larger than the heating rates, 
whereas the opposite orange region shows that the heating rates are larger rate than cooling rates.
The solutions with positive slope, i.e, $dT/d\Xi >0$ are thermally stable because 
if the temperature rises slightly from this curve fixing $\Xi$, the cooling rates are 
larger than the heating rates and the temperature goes down.
For the opposite reason, those with negative slope are thermally unstable. 
The changing points of the sign of slope are instability points.

 This thermal equilibrium curve has a very complex shape, with four stable branches separated
by regions of instability and/or rapid change (Fig.\ref{fig:continuum}b, with zoom in
Fig.\ref{fig:continuum}c).  This is very different to the single S shaped thermal
equilibrium curve using in B83, where there is a minimum pressure
ionisation parameter associated with the material on the hot branch,
$\Xi_\mathrm{h,min}$ at the temperature $\sim \frac{1}{2}T_\mathrm{IC}$ which marks the
bottom of the heated atmosphere, while the maximum pressure ionisation
parameter associated with material on the cold branch $\Xi_\mathrm{c,min}$,
marks the top of the disc. Instead, for this more complex shape, we
take the minimum pressure ionisation parameter of the long middle branch
(associated with partially ionised Oxygen and Iron L shell) to mark the bottom
of the heated atmosphere at i.e. we set $\Xi_\mathrm{H,min}=\Xi_\mathrm{m,min}=8.2$, 
where $T\ll T_\mathrm{IC}$. 

Fig.\ref{fig:force m}a shows this equilibrium curve in the
standard ionisation parameter representation. The $\Xi_\mathrm{m,min}$ point
corresponds to material with $\xi=170$, where the soft X-ray opacity
is already substantial.  The maximum pressure ionisation parameter on
the cold branch, $\Xi_\mathrm{c,max} \sim 7.4$ has standard $\xi\sim 14$ (see
Fig.\ref{fig:force m}a), so it is underneath layers with $14\le \xi\le 170$. This
material has very substantial opacity so it seems very unlikely that
it can be irradiated directly by the source SED. Hence we assume the
disc surface also has $\Xi_\mathrm{m,min}$ i.e. we set
$\Xi_\mathrm{C,max}=\Xi_\mathrm{m,min}$. These are almost identical in $\Xi$ (see
Fig.\ref{fig:continuum}c), but quite different in $\xi$ due to their very different
temperatures. We use thermal equilibrium to set density at the 
disc surface to $\xi =170$. At each time step, the density of disc surface 
is update via $n =L_x/(170 R^2)$ and check that temperature  is  
hotter than that of the viscous heated disc across the entire grid.

\subsection{Inner attenuating corona}
\label{sec:inner_corona}

\begin{figure}
    \centering
    \includegraphics[width=\hsize]{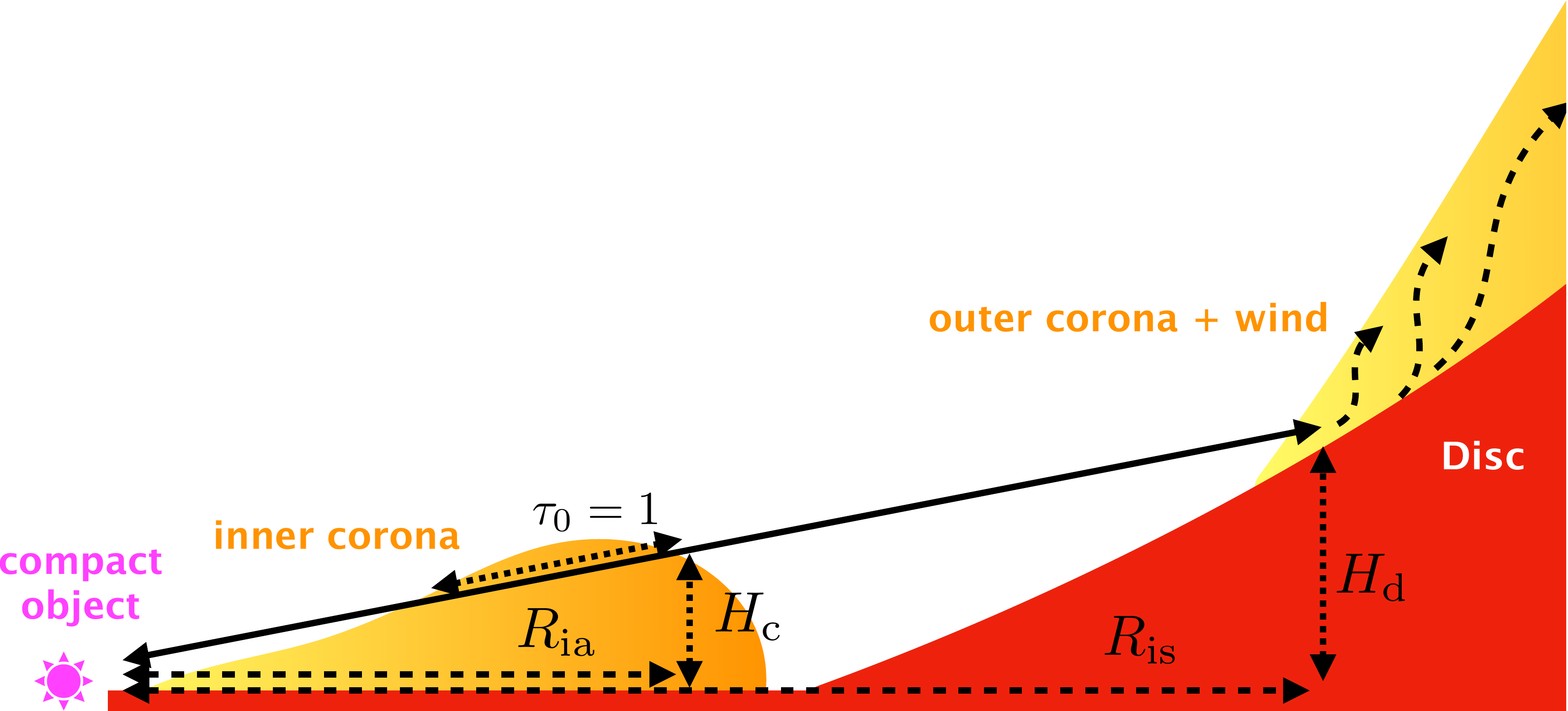}
    \caption{schematic view of an inner corona and a wind}
    \label{fig:corona}
\end{figure}

The original thermal wind paper of B83
gives an overview of the structure of the Compton heated upper
layers of the disc for optically thin material. In this limit, the
heated material above the inner disc forms a static atmosphere with
scale height $H_c/R\sim (v_\mathrm{IC}/v_g)=(T_\mathrm{IC}/T_g)^{1/2}$ where $v_g$
($T_g$) is the escape velocity (virial temperature). This gives
$H_c\sim [2R^3/R_\mathrm{IC}]^{1/2}$.  However, \citet[hereafter BM83]{Begelman1983b} 
shows that it is easy for this heated atmosphere to go
optically thick in the radial direction along the disc plane, so it
forms an inner attenuation zone, reaching $\tau=1$ at $R_\mathrm{ia}$ given
by
\begin{equation}
\frac{R_\mathrm{ia}}{R_\mathrm{IC}}= 0.021\Bigl[ \frac{T_\mathrm{IC,8} (L/L_\mathrm{Edd}) }{\Xi_\mathrm{H,min} } \Bigr]^{1/2}
\end{equation}
For our soft state simulation this gives $R_\mathrm{ia}\sim
0.0004R_\mathrm{IC}\sim 200R_g$. This radius is on scales which are more than an
order of magnitude smaller than the starting point of our radial grid,
so this structure cannot be resolved by our simulation. However, it
strongly affects the wind properties in the simulation range as it
casts a shadow over the disc surface out to large radii (see Fig.\ref{fig:corona}).
The outer disc is only directly irradiated again when the disc
scale height increases by a large enough factor that it rises above
the shadow zone. For an irradiated disc with scale height given by
equation 4, then the shadow cast by the inner attenuation zone ends
at $R_\mathrm{is}$, where the disc is directly illuminated again. The
geometry gives $H_d/R_\mathrm{is}=H_c/R_\mathrm{ia}$ so that
\begin{equation}
R_\mathrm{\mathrm{is}}= 3.0\times 10^7 ~T_\mathrm{IC,8}^{7/8}~ \left( \frac{M_c}{M_\odot} \right)^{1/2} \left( \frac{L}{L_\mathrm{Edd}} \right)^{3/8} \Xi_\mathrm{h,min}^{-7/8} R_\mathrm{g}
\end{equation}

This gives $R_\mathrm{is}=0.18R_\mathrm{IC}$ for $L/L_\mathrm{Edd}=0.3$ for our assumed system parameters with the soft state SED. 
We incorporate the sub-grid physics of this inner attenuating corona by
changing the illumination pattern onto the disc in the outer regions
which are covered by the hydrodynamic grid. We define a critical angle
from the mid-plane $ \alpha_c = H_c/R_\text{ia}$ and assume that
the optical depth from the centre to any point on the disc surface has
$\tau=\exp[1-(\alpha/\alpha_c)^2]$ and use this to attenuate the X-ray
luminosity by $e^{-\tau}$ before it enters the grid.

\section{results for fiducial soft state spectral energy distribution}

\begin{table*}
    \centering
    \caption{The summary of simulations and inner corona parameters}
    \begin{tabular}{c|c|c|c|c|c|c|c|c|c|c|c|c}
    \hline
   SED$^a$ & $L/L_\mathrm{Edd}$ &$F_\mathrm{rad}$ & $T_\mathrm{IC}~[10^8 \mathrm{K}]$& $R_\mathrm{IC}/R_\mathrm{out}$&$H_\mathrm{out}/R_\mathrm{out}$&$\Xi_\mathrm{h,min}$&$R_\mathrm{ia}~[R_\mathrm{g}] $&$H_\mathrm{c} ~[R_\mathrm{g}]$&$R_\mathrm{is}/R_\mathrm{out}$ &$R_\text{disc}/R_\text{out} $& $\dot{M}_\mathrm{w} ~[10^{18}~\mathrm{g~s^{-1}}]$ &$\dot{M}_\mathrm{w}/\dot{M}_\mathrm{a} ^b$\\
    \hline
    S & 0.3  & Y  & 0.10 & 1    & 0.044 & 100 &240 &6.4 &0.18 &1.0 & 21  & 6.0\\
    S & 0.3  & N  & 0.10 & 1    & 0.044 & 100 &240 &6.4 &0.18 &1.0 & 8.7  & 2.5\\
    S & 0.3  & Y  & 0.10 & 1    & 0.044 & 100 &240 &6.4 &0.18 &0.18 & 1.2  & 0.34\\
    S & 0.5  & Y  & 0.10 & 1    & 0.047 & 100 &310 &9.5 &0.22 &1.0 & 28  & 4.8\\
    S & 0.5  & N  & 0.10 & 1    & 0.047 & 100 &310 &9.5 &0.22 &1.0 & 14  & 2.4\\
    S & 0.1  & Y  & 0.10 & 1    & 0.038 & 100 &140 &2.8 &0.12 &1.0 & 1.8 & 1.5 \\
    S & 0.3  & Y  & 0.10 & 1    & 0.044 & --  &640 &28  &1.0  &1.0  & 5.6 & 1.6\\ 
    S & 0.5  & Y  & 0.10 & 1    & 0.047 & --  &730 &34  &1.0  &1.0  & 7.9 & 1.3\\
    H  & 0.06 & N & 0.70 & 0.14 & 0.035 & 3.3 &220 &22  &11   &1.0   & 0.43  &0.61\\
    \hline 
    \multicolumn{12}{l}{$^a$ S:soft state, H:hard state}\\
    \multicolumn{12}{l}{$^b~\dot{M}_a = L/(0.1c^2)$ }
    \end{tabular}
    \label{tab:summary}
\end{table*}
We run a series of simulations as described below, with parameters and resulting mass loss rates 
given in Tab.\ref{tab:summary}. 
All simulations run for 9 sounds crossing time ($c_\mathrm{IC}/R_{IC}=\sqrt{k T_\mathrm{IC}/(\mu m_p)}/R_\mathrm{IC}$). 
This correspons to $1.7\times 10^5~\mathrm{s}$.

\subsection{Fiducial Simulation: effect of radiation force}
\label{sec:fiducial}

\begin{figure}
\includegraphics[width=0.9\hsize]{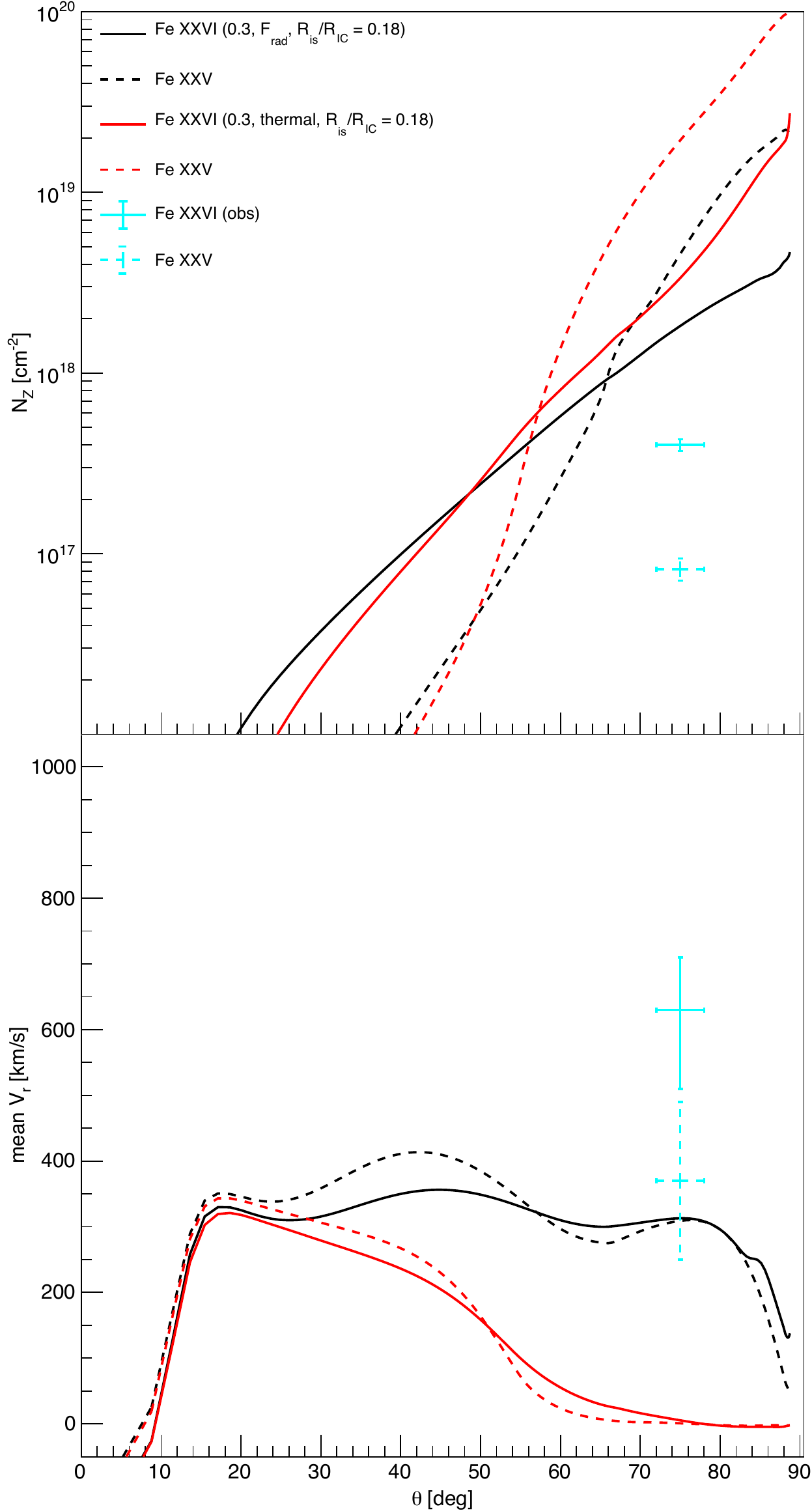}
\caption{Top : the inclination dependence of column densities of Fe {\scriptsize XXVI} (solid line) and Fe {\scriptsize XXV} (dashed line).  
Bottom: the the inclination dependence of column weighted mean velocities of Fe {\scriptsize XXVI} (solid line) and Fe {\scriptsize XXV} (dashed line). 
The colours show the thermal-radiative wind of $L/L_\mathrm{Edd}=0.3,~R_\mathrm{IS}/R_\mathrm{IC}=0.18$(black), the thermal wind (red).
%Columns are larger than observation. The column of Fe {\scriptsize XXVI} are smaller than that of Fe {\scriptsize XXV} in each simulation but observational values of these are opposite. This means the ionisation state of each simulation is lower than observational one. 
}
\label{fig:integral_03Ledd}
\end{figure}

We first run a fiducial simulation, for the soft state SED in Fig.\ref{fig:continuum},
assuming $L/L_\mathrm{Edd}=0.3$. We include all radiation force terms
(electron scattering, bound-free and line), attenuation of these effects and fiducial inner corona defined by Eq.\ref{eq:tau}.
The derived density and temperature structure are shown in the Appendix
(top left in Fig.\ref{fig:temperature}) and the total mass loss rate through
the outer boundary is $21\times 10^{18}$~g~s$^{-1}$ (see Table 1).
The black lines in Fig.\ref{fig:integral_03Ledd}
show the column density (upper panel) and velocity (lower panel) as 
a function of inclination angle for this simulation, with Fe {\scriptsize XXVI} as the solid line
and Fe {\scriptsize XXV} as dashed. 

We compare this to the results from a 
simulation where we turn off all the radiation force terms 
and their attenuation except inner corona ($M=0$ in Eq.\ref{eq:tau}), 
so that the wind only is thermally driven (top middle in Fig. \ref{fig:temperature} and red lines in Fig. \ref{fig:integral_03Ledd}). 
This simulation has much lower total mass loss rate, of $8.7\times 10^{18}$~g~s$^{-1}$, 
so it is clear that radiation force is important, 
and that this wind is better described as thermal-radiative
rather than simply thermal. 
The thermal wind has a very low velocity at high inclination angles, close to the equatorial plane of the disc, 
so it has very large column density in these directions. 
Neither of these matches well to the observations
at the inferred high inclination angle of H~1743-322 (shown by the cyan points). 

Including the full radiation force terms gives a  dramatic increase in velocity at large inclination angles.
This is because this material close to the disc is mainly on the middle branch of the thermal equilibrium curve i.e. is at an ionisation parameter close to $\xi\sim 170$, 
with the corresponding temperature which is substantially less than the Compton temperature, 
so the material forms a mainly static atmosphere rather than an outflowing corona
\citep{Higginbottom2015,Higginbottom2018}. The presence of the middle branch of the S curve for this soft continuum spectrum gives the difference between these calculations and the earlier exploration of \citet{Proga2002}, where they showed that the radiative force was negligible for a much harder continuum. 

Instead, when radiation force is included, 
this warm, partially ionised material has enough opacity for bound-free and bound-bound opacity to accelerate it out so it can escape as a wind. 
This increases in velocity more than offsets the increased amount of material which escapes, so the column density
decreases. 

We investigate which term of the radiation force is most important at any $\xi$. 
Fig.\ref{fig:force m}b shows force multiplier of the continuum process (red),
lines process (green) and their sum (black) at thermal equilibrium state. 
Radiation force on free elections may have some impact on the simulations, 
but at our luminosity of only $0.3L_\mathrm{Edd}$, this alone is not enough to unbind much material. 
We run an additional simulation including only this term, and find a total
mass loss rate of $8\times 10^{18}$ g~s$^{-1}$, so same within 10\%.
than the purely thermal wind. Fig.\ref{fig:force m}b shows that the radiation force
on lines and photo-electric opacities have similar magnitudes, and that
adding all the terms together gives a force multiplier of around $M
= 9.6$ at $\xi = 170$, thus the ratio of total radiation force to gravity is $M\times
L/L_\mathrm{Edd}=0.3\times 9.6 > 1 $. Neglecting line opacity but
including electron scattering and bound-free gives a force multiplier
of $3.6$, so this is just enough to get to unbind the material when it
is launched, but not enough to continue accelerating it once it
becomes more ionised. We rerun a simulation including only electron scattering and bound-free opacities
and find a mass loss rate of $9.8\times 10^{18}$ g~s$^{-1}$. Thus it is the combination of all
opacity sources which is important at high but subcritical $L/L_\mathrm{Edd}$
values, and a single correction factor for $L/L_\mathrm{Edd}$ (as in D18) is
too simplistic to describe the behaviour revealed 

\subsection{Changing the disc size}
\label{sec:ch_disc}
\begin{figure}
    \centering
    \includegraphics[width=0.9\hsize]{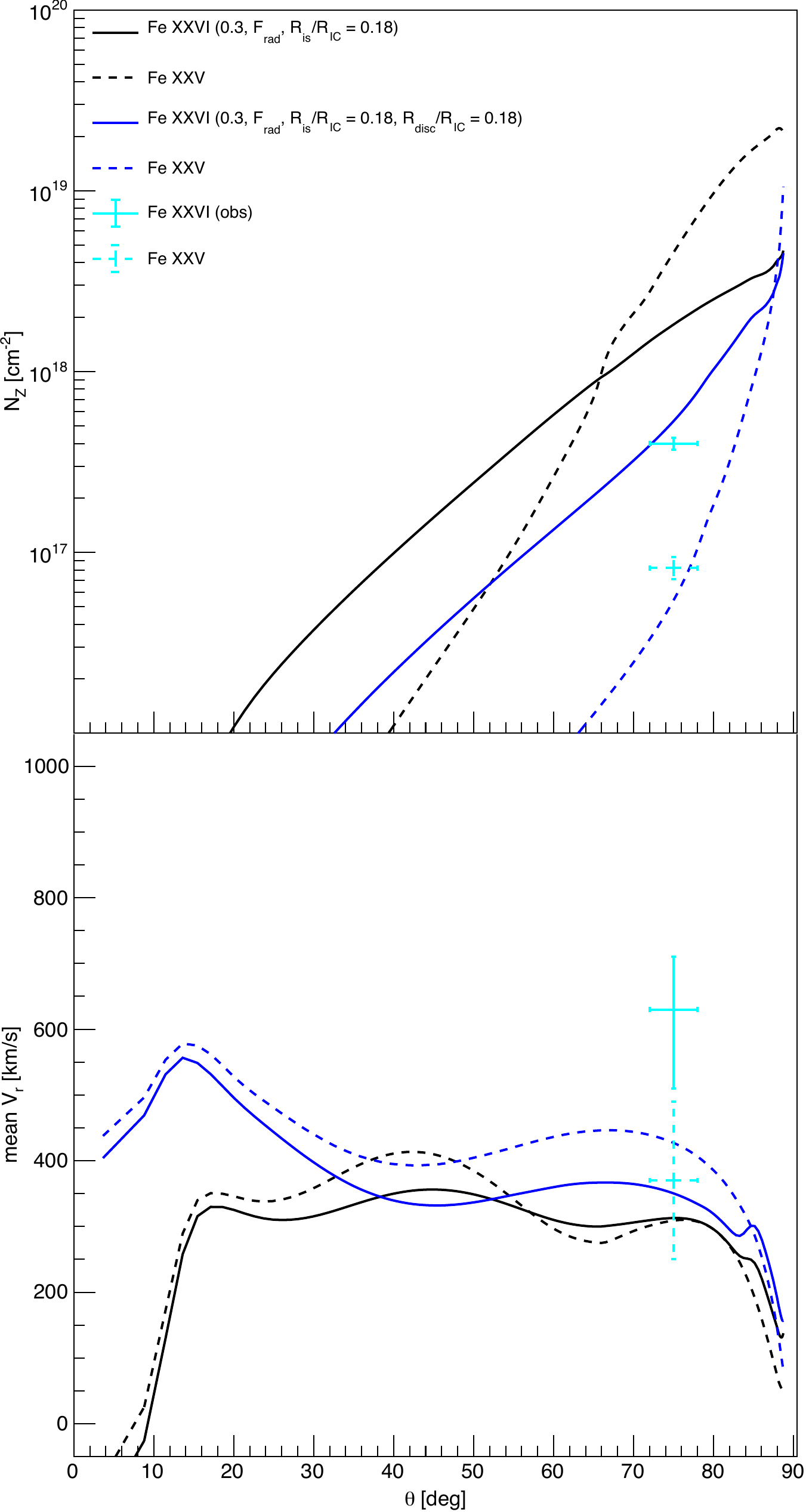}
    \caption{As in Fig. \ref{fig:integral_03Ledd} but $R_\text{disc}/R_\text{IC}$ = 1.0 (black) and 0.18 (blue). }
    \label{fig:integral_small_disc}
\end{figure}

Thermal winds depend most sensitively on the shape of the spectrum, the luminosity, and the size of the disc. The spectral shape is defined by observations, but the overall luminosity and disc size depend on the assumed system parameters which are poorly known. The analytic models give a dependence on $N_H\propto \log R_\text{disc}/R_\text{in}$ where $R_\text{in}=0.2 R_\text{IC}$ is the launch radius of the wind. However, the reduction in column required from the data is
not simple to produce in the analytic approximation of D18 as Fe {\scriptsize XXVI} should decrease by factor 5 but that of Fe {\scriptsize XXV} should decrease by 2 orders magnitude at $75^\circ$. However, we note that our disc size is already only $R_\text{disc}=R_\text{IC}$, and the analytic approximations probably become unreliable as we approach the wind launch radius. 

Instead, we reduce the disc size to $R_\text{disc}=0.18 R_\text{out}$,  which is the same radius as inner shadow radius as well as the wind launch radius, then we re-run the simulation including full radiation force. 
We run the simulation over the same grid as before, but set a very low density at mid-plain ($\theta = 90^\circ$) when the radius is larger than the disc radius. 
The blue line in Fig. \ref{fig:integral_small_disc} shows the results of this simulation.
We can reproduce the ion columns seen in the data, and in particular we now have a higher column of
Fe {\scriptsize XXVI} than Fe {\scriptsize XXV}, indicating a higher ionisation state for the wind. 
The wind is launched, but only just! 
The mass loss rate is 20 times smaller than that of fiducial simulation, so the density of the wind is  quite low. 
It is still slightly overshadowed, so the disc is not fully illuminated by the central source, but the drop in overall mass accretion rate reduces the density sufficiently for the ionisation parameter to be higher than before. 
This means that the contribution of line and photo-electric absorption to the driving is almost negligible. 
This wind is driven by the combination of radiation force on electrons and gas pressure gradient force, and its velocity matches well to that measured in the data. 

We explore the effect of some of the other parameters input into the fiducial simulation below in order to illustrate their impact on the predicted wind, but this is our best match overall to the observed data. 

\subsection{Changing luminosity:$L/L_\text{Edd}=0.5$ and $0.1$}
\label{sec:ch_luminosity}
\begin{figure}
\includegraphics[width=0.9\hsize]{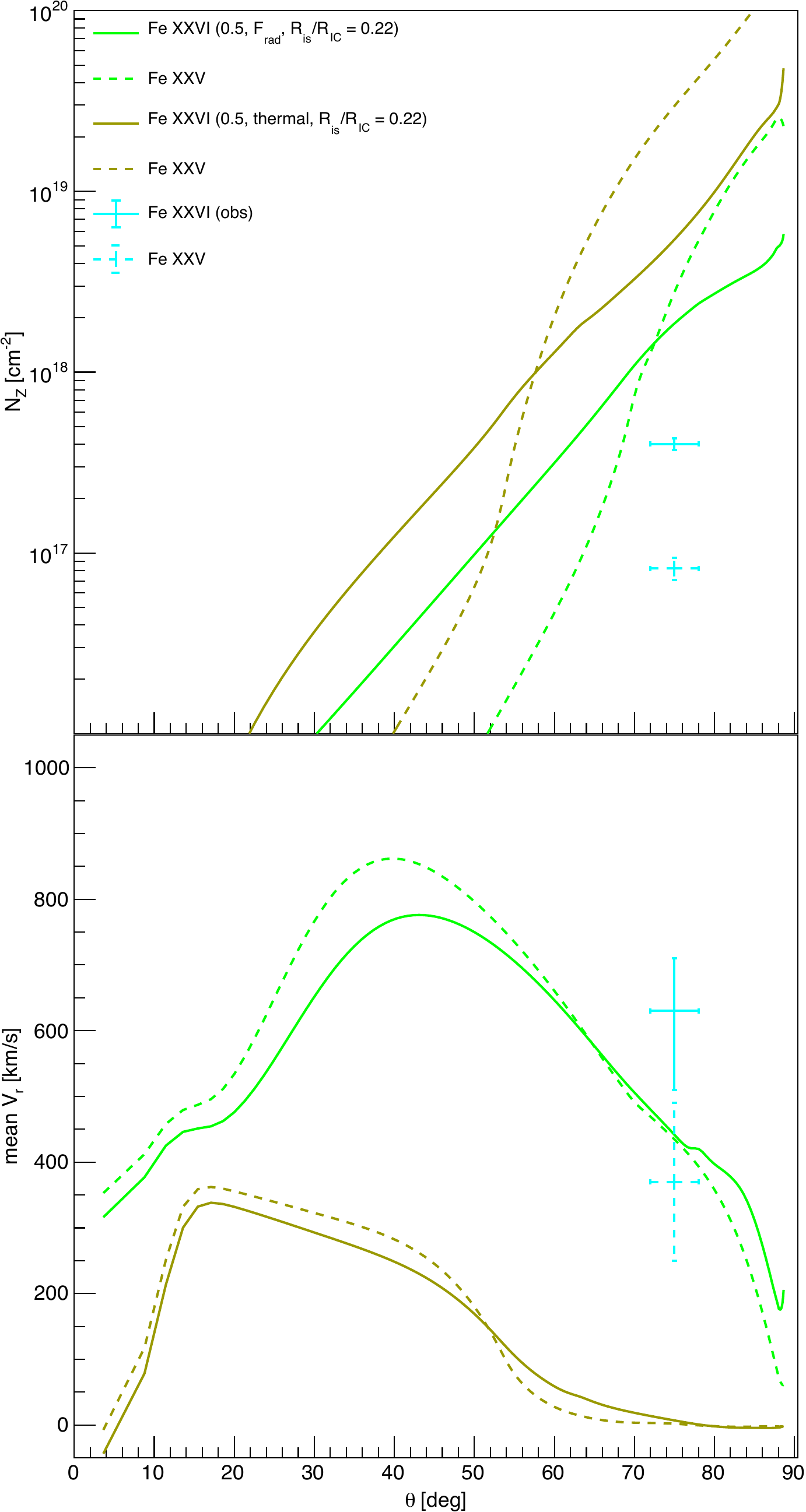}
\caption{As Fig.\ref{fig:integral_03Ledd} but for $L=0.5 L_\mathrm{Ldd}$.}
\label{fig:integral_05Ledd}
\end{figure}

\begin{figure}
\includegraphics[width=0.9\hsize]{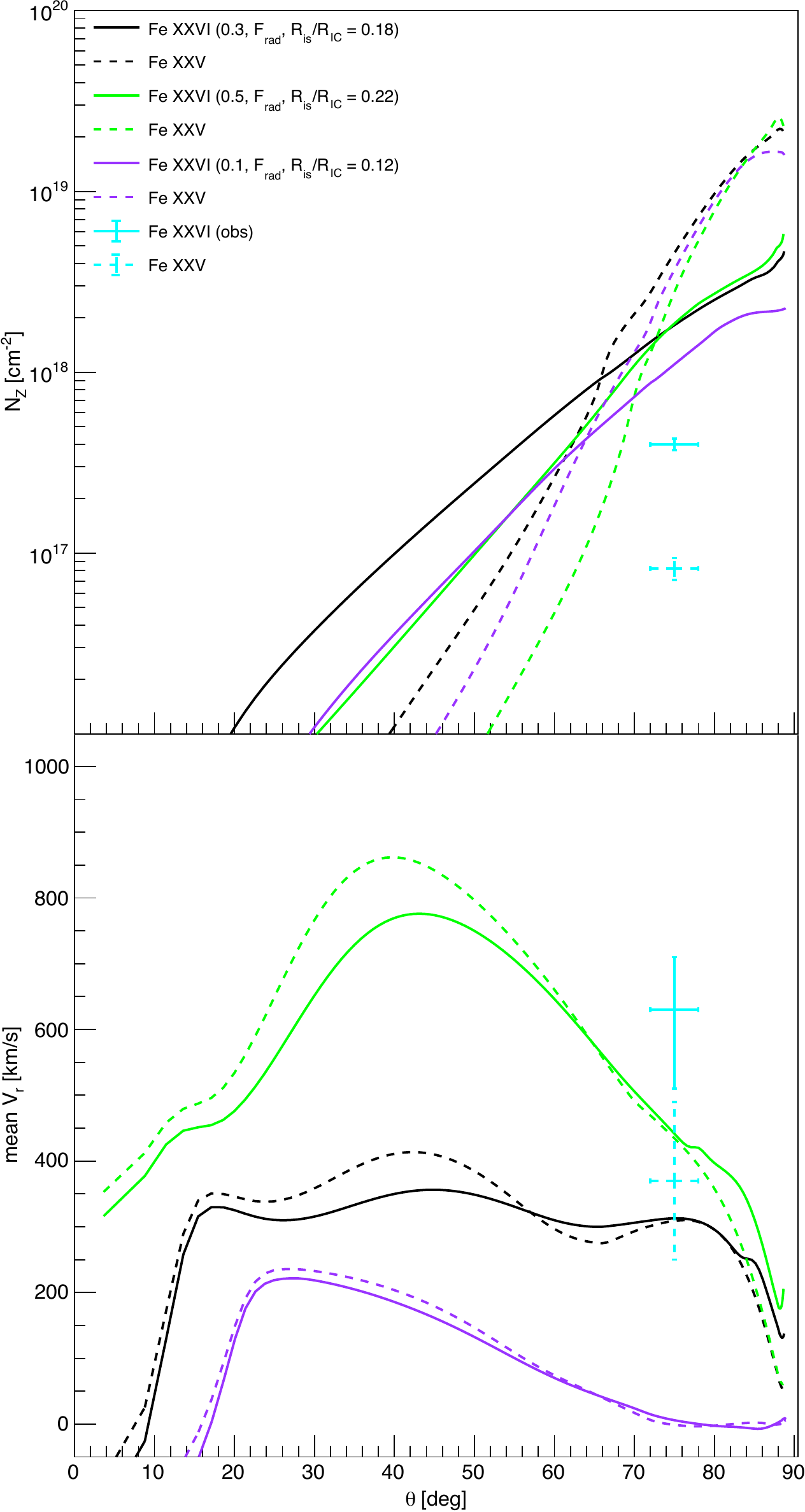}
\caption{As in Fig.\ref{fig:integral_03Ledd} but for $L/L_\mathrm{Edd} = 0.1~(\mathrm{violet}), 0.3~(\mathrm{black}) ,0.5~(\mathrm{green})$ }
\label{fig:integral_3}
\end{figure}

We now consider the effect of changing luminosity.  We run an
additional set of simulations with $L/L_\mathrm{Edd}=0.5$ for the same SED. 
This changes the radius of inner attenuation zone and the outer disc scale height, so that the shadow extends to  $R_\mathrm{is}=0.22R_\mathrm{out}$. 
Fig.~\ref{fig:integral_05Ledd} shows the resultant column density (upper) and velocity (lower)
for this simulation including all radiation forces (green) compared to a purely thermal wind at the same
luminosity (yellow). 
The purely thermal results are very similar to those at $L/L_\mathrm{Edd}=0.3$, with very low velocity at high inclination angles, and consequently large column density. 
This material is again mainly on the middle branch of the ionisation instability, 
so it does not have high enough temperature to escape. 
However, including radiation force on this material makes even more difference at these higher luminosities,
so the column is lower and the velocity higher than for
$L/L_\mathrm{Edd}=0.3$, bringing the models closer to the observed data points (cyan).

We also run a simulation including all radiation force terms  with the same SED but $L/L_\mathrm{Edd}=0.1$. 
Fig.\ref{fig:integral_3} shows the column and velocity from this (violet) compared to $L/L_\mathrm{Edd}=0.3$ (black)
and $0.5$ (green) including all radiation force terms. 
The column densities in each ion are remarkably similar,  but the velocity increases dramatically at high inclination angles, and the total mass loss rate increases from
$1.8$ to $21$ to $28\times 10^{18}$~g~s$^{-1}$. 

Fig.\ref{fig:integral_3} clearly shows that thermal-radiative winds are fast enough to match the observations for $L/L_\mathrm{Edd}\gtrsim 0.3$, but that these luminosities give columns which are a factor 3--10 larger than observed. 
There are multiple ways we can reduce the wind efficiency.
As we already indicated in Sec.\ref{sec:ch_disc}, the first is  by reducing the size or height of the outer disc (the system parameters are quite uncertain, and the disc scale height need not exactly follow the irradiated disc shape (see e.g. \citealt{Kimura2019}), the second is by changing the irradiation pattern as a function of angle. 
Section 3.3 assumed a very simple exponential attenuation with angle produced by the inner corona.
Density structure in the inner corona could  give slightly stronger attenuation of both the disc and X-ray source at large inclination angles, 
or the different radiation pattern of the flat disc and more isotropic X-ray source could  give different spectral illumination of the disc surface compared to that observed. 

\subsection{Changing the extent of the shadow from the inner attenuation zone}
\label{sec:ch corona}

\begin{figure}
\includegraphics[width=0.9\hsize]{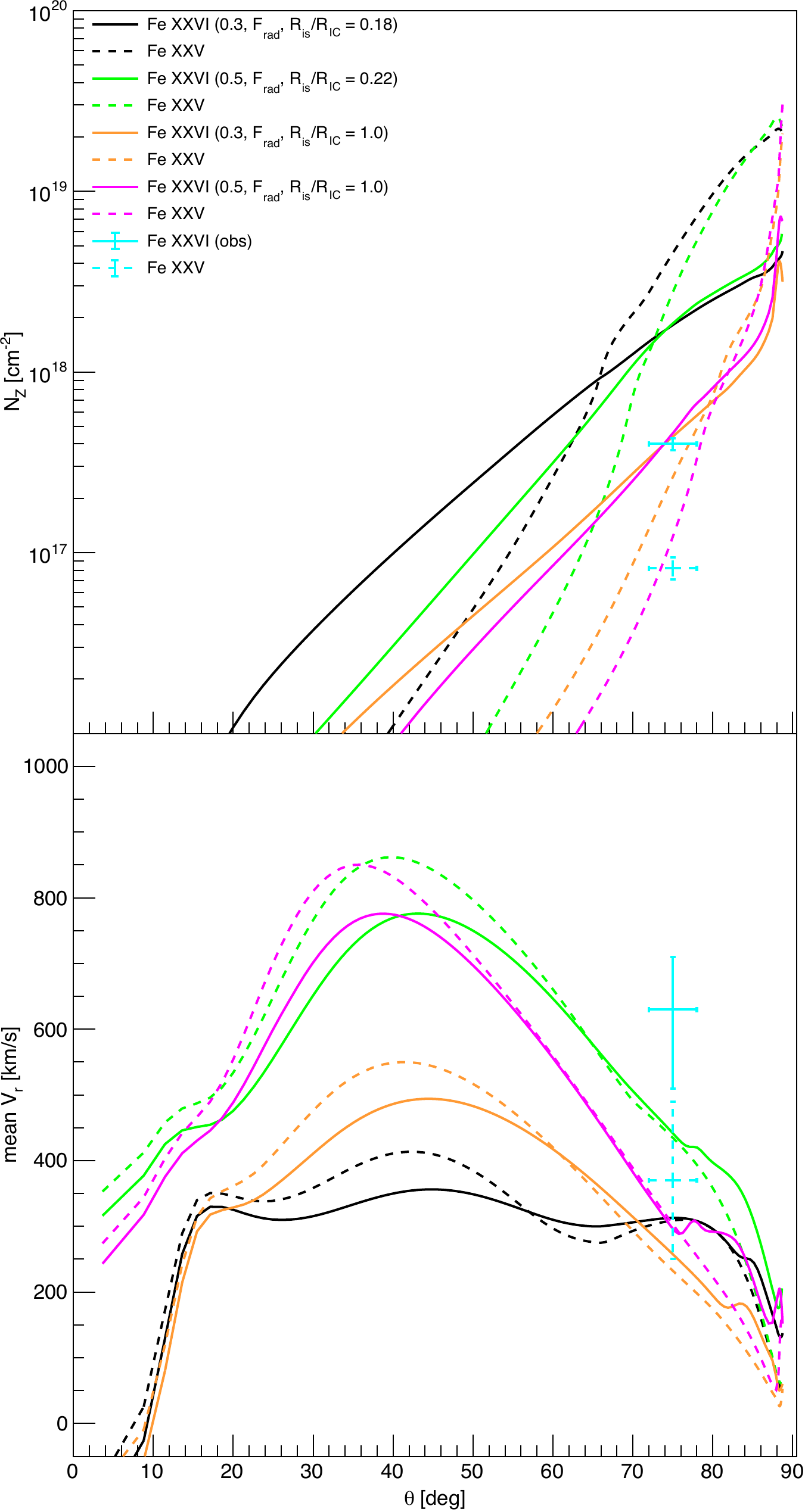}
\caption{As in Fig.\ref{fig:integral_03Ledd} but for
$L/L_\text{Edd} = 0.3,~ R_\text{IS}/R_\text{IC} = 0.18~(\text{black}), ~
L/L_\text{Edd} = 0.5,~R_\text{IS}/R_\text{IC} = 0.22~(\text{green}),~
L/L_\text{Edd} = 0.3,~R_\text{IS}/R_\text{IC} = 1.0~(\text{orange}),~
\text{and}~ L/L_\text{Edd} = 0.5,~R_\text{IS}/R_\text{IC} = 1.0~(\text{magenta})$.}
\label{fig:integral_4}
\end{figure}

We quantify the ideas above by simply reduce the illumination of the outer disc by 
changing the shadow size to
$R_\text{is}/R_\mathrm{IC}=1.0$ on each of the simulations for $L/L_\mathrm{Edd}=0.3$ and $0.5$,
including full radiation force. 

The lines in black and orange on
Fig.~\ref{fig:integral_4} shows the effect of this for
$L/L_\mathrm{Edd}=0.3$, while the lines in green and magenta show this for 
$L/L_\mathrm{Edd}=0.5$. The extended shadow means that the illuminating flux is 
lower for the given simulation, which means that the wind mass loss rate 
is lower (D18). However, the velocity remains mainly unchanged, as once the wind rises up it 
sees the same radiation force as before. Hence this makes the 
column lower, whilst maintaining the fast velocity, giving a better match to the data (cyan). 

A changing radiation pattern from the flat disc and more isotropic X-ray source would have a similar effect in 
reducing the illumination of the disc surface, but keeping the full radiation force on the wind once it rises up,
though this is more complex to model as the Compton temperature would also change as a function of height.

\section{Result for the hard state spectral energy distribution}
\label{sec:hard state}
\begin{figure}
    \centering
    \includegraphics[width=0.9\hsize]{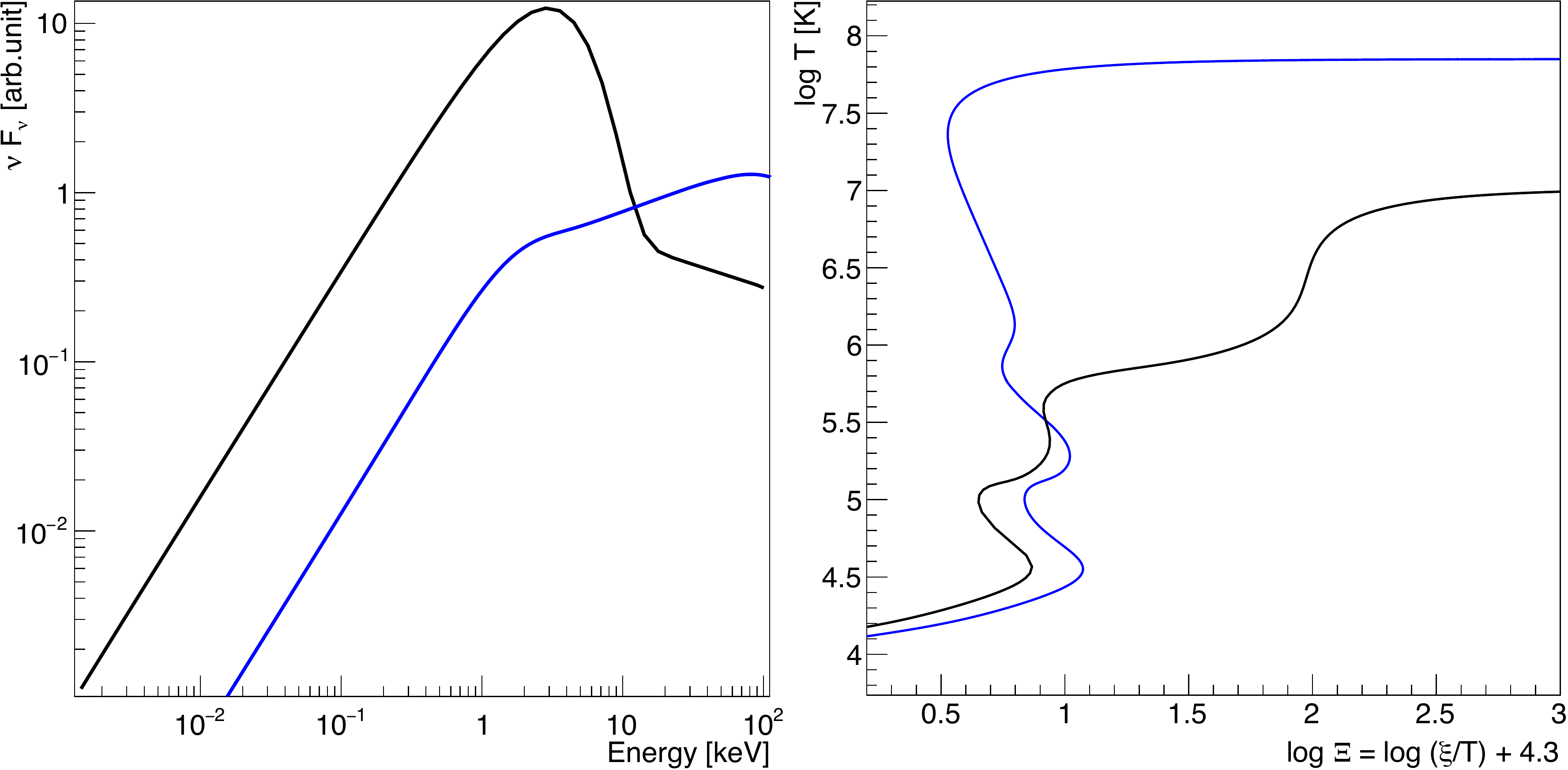}
    \caption{a) Spectra of soft state(black) and hard state (blue). 
    b) The thermal equilibrium curves. colours correspond to input spectra of a) .
    }
    \label{fig:thermal_eq_sp}
\end{figure}
\begin{figure}
    \centering
    \includegraphics[width=0.9\hsize]{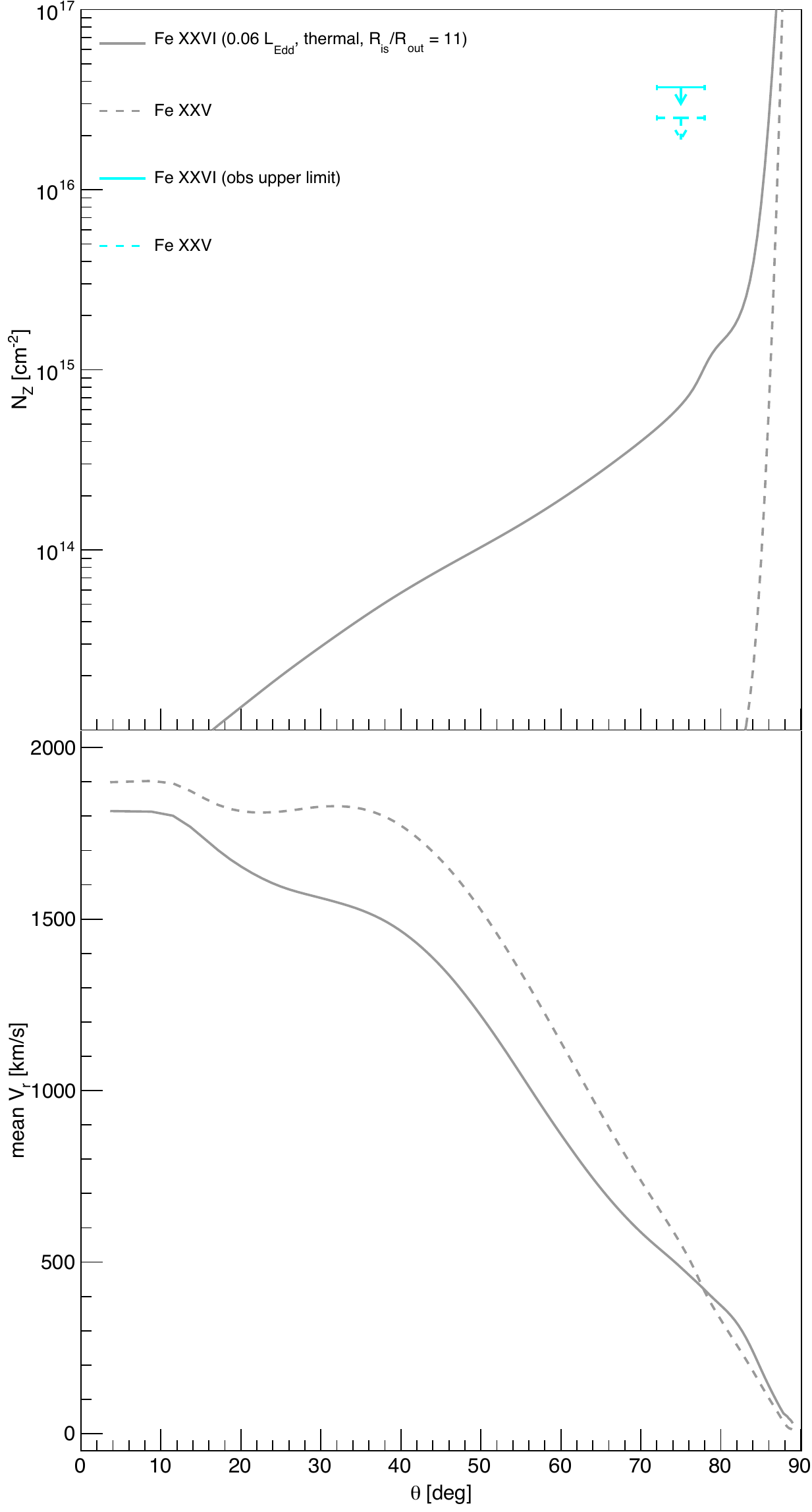}
    \caption{The angular dependence of column densities Fe {\scriptsize XXVI} (black solid line), Fe {\scriptsize XXV} (black dashed line) 
    and those of column weighted velocites (bottom) of hard state. 
    The cyan colars show the observational upper limit calculated from \citet{Miller2012}}
    \label{fig:column_hard}
\end{figure}
We now consider the predicted wind properties in the hard spectral state. 
The yellow line in Fig. \ref{fig:thermal_eq_sp}a shows the observed SED after the switch to the hard state. 
This has a much higher Compton temperature than the soft state (black), 
and the thermal equilibrium curve is now closer to the classic S curve of B83.
The luminosity is fairly low,
but the high Compton temperature means that the thermal wind can be launched from closer to the black hole than in the soft state, 
so it can be more highly ionised (see D18).
However, the key new aspect in this work is the realisation that the inner attenuation zone also responds to the changing SED.
The higher Compton temperature means that this has higher scale height, so it now casts a deep shadow over the entire outer disc, with $R_\text{is} = 11 R_\text{out}$.
This completely suppresses the thermal wind unless the hard X-ray source has scale height which is larger than the inner attenuation zone $H_\text{c}\sim 20 R_\text{g}$.
The response of the wind to the changing SED is not then simply due to the wind being launched from closer in,
and being more highly ionised, as suggested by D18.
Instead, here there is no thermal wind since the outer disc is not irradiated.
The inner attenuation zone responds to the changing SED,
increasing in scale height and hence increasing the shadow cast over the disc. 

We run a final simulation for this hard state to confirm that the wind is 
suppressed sufficiently to match with the observational data. We use the hard state spectrum to calculate
a new set of heating and cooling rates. These define the disc surface at $\Xi_\text{c,max}=12, ~T_\text{c,max}=3.5\times10^4 \mathrm{K}$ which gives the $\xi_\text{c,max}=22$ from thermal equilibrium curve (blue line in Fig.\ref{fig:thermal_eq_sp}b). Thus, we set the density at the disc surface as $n=L_x/(22R^2)$ at each time step and the grid size is same as Sec.4. We include the illumination attenuation from the 
inner corona but do not include line force as this is entirely negligible for highly ionised, very optically thin material. 

Fig. \ref{fig:column_hard} shows that the predicted 
iron columns are more than a factor 10 lower than the observational upper limits (the density and temperature structure are Fig. \ref{fig:map hard-state}), 
so our thermal wind simulation can indeed explain the disappearance of the wind in the hard state.
This is because of over ionisation by the hard X-ray and large size of inner corona. 
In the hard state, when the thermal instability once occurs, the temperature rapidly heats up to the Compton temperature $T_\text{IC} = 7.0\times 10^{7} \text{K}$ (Fig. \ref{fig:map hard-state}).
This rapid rise gives strong gas pressure gradient force, so velocity is lager than that of soft state (Fig.\ref{fig:column_hard}b).
The large velocity lead the low density if we consider mass continuity, also the large inner corona suppress  the wind density.

However, there is still a wind, and the mass loss rate is over half of the mass accretion rate. 
We can understand this result by using the analytic estimates in D18.
These show that the critical luminosity, where the Compton heating is sufficient to produce a wind from $R_\text{IC}$, is $L_\text{crit}\sim 0.04$. 
This is below the observed luminosity of $L_\text{bol}=0.06~L_\text{Edd}$, 
but the attenuation by the inner corona means that the flux illuminating the inner edge of the hydrodynamic grid is a factor 10 lower.
Thus $L/L_\text{crit}\sim 0.1$, but the disc is large compared to
$R_\text{IC}$ ($R_\text{out}/R_\text{IC} = 7.0$),
so the wind is still quite effective (see cyan line in Fig 3 of D18). 
Thermal winds have a large impact on the mass available for the 
outburst for systems with a large disc, explored in more detail in Dubus et al (2019, in prep).

\section{Discussion}
We have shown results from the first radiation hydrodynamic simulations of a thermal-radiative wind.
Including the radiation force gives important differences to the structure of the thermal wind for $L > 0.1L_\text{Edd}$ (Sec.\ref{sec:fiducial} ,\ref{sec:ch_luminosity}).
We include electron scattering, bound-free absorption and line opacity as these all give corrections which are of similar order,
and all of them together act to produce a force multiplier which is large enough for these sub-Eddington flows to become super-Eddington.
Thermal winds alone are too slow to match to the observed winds (see also \citealt{Higginbottom2018}),
but these thermal-radiative winds can produce both the column density and velocity of the material seen in the soft state of H1743-322. 

We identify the key role played by the inner static corona,
on size scales of a few hundred $R_\text{g}$ (Sec.\ref{sec:ch corona}, ).
This material is part of the Compton heated atmosphere of the disc,
but is on size scales much less than $R_\text{IC}$ so it is bound, 
and has $H\ll R$.
Nonetheless, it is important in determining the illumination of the outer disc where the wind arises. 
It is very easy for this corona to become optically thick in the equatorial plane, 
so that it shadows much of the outer disc from the inner disc emission. 
This material is below our grid scale, so we incorporate it analytically following BM83. 
The scale height of the inner corona casts a shadow which prevents direct illumination of the outer disc until it rises out of this shaded zone because of the intrinsically concave (saucer-like) shape of the disc. 
The small scale height of the inner attenuating corona in the soft state means that the outer disc is directly illuminated in the soft state. 
Conversely, the much larger scale height of the inner corona in the hard state completely shadows the entire disc, so the wind is suppressed.  
This contrasts with the explanation in D18 where the thermal wind is still present in the hard state, 
but is less visible due to its higher ionisation which is a consequence of its smaller launch radius (Sec.\ref{sec:hard state}). 
Reality may be a mixture of the two, as the structure of the hard X-ray source itself is changing during the transition.
The inner disc evaporates into an X-ray hot flow whose scale height probably increases as the source dims.
Evidence for this is that the electron temperature increases as shown by \citet{Motta2010}, 
which can be modelled by an expanding hot flow region \citep{Gardner2014, Kara2019,Marcel2018b}.
The disc only starts further out, so the inner shadow corona starts further out, and the larger scale height of the X-rays means that some fraction can directly illuminate the outer disc as they extend above the shadow corona. 
A change in illumination pattern should have observable consequences on the optical continuum as well as on the X-ray wind properties.
In bright BHB the optical continuum is generally due to reprocessing of X-rays in the outer disc, showing that the outer disc is illuminated \citep{vanParadijs1996}.
This reprocessed optical emission can be seen directly in broadband spectra in the soft state \citep{Hynes2002,Kimura2019},
and especially in fast variability, where the optical emission is variable on timescales of $\sim 1-20$~s, lagged behind but correlated with the X-rays \citep{O'Brien2002}.
The hard state is more complex, but close to the spectral transition there are still clear signatures that part of the optical emission is produced by reprocessing in the outer disc (\citealt{Hynes2009, Veledina2017} for Swift J1753, though this reprocessing signal disappears as the source dims:\citealt{Veledina2017}).
Simultaneous fast optical and X-ray variability data can directly measure the changing irradiation pattern on the disc and test these models of an inner attenuation zone. 

The inner corona can also have an impact on the soft state spectrum.
The shadow affects the corona structure when it becomes optically thick along the disc direction, but this is also associated with a vertical optical depth which is $\sim 0.05$.
Thus 5\% of the inner disc flux is scattered rather than being directly produced in the disc photosphere,
and so at this level, it is clear that the disc emission should differ from even the best pure disc photosphere calculation such as those of \citet{Davis2005, Davis2006a}.
This high ionisation state layer on top of the disc will also change the reflected coronal flux from the disc,
with around 5\% of the corona flux being scattered from this much higher ionisation state layer than from the disc itself. 

\section{Conclusions}
We show that thermal-radiative winds can match the observed wind properties in H1743-322. 
This includes both the column density and velocity of the wind seen in the soft state.
We confirm the results of \citet{Higginbottom2018} that thermal winds alone are too slow, but we show that radiation pressure (on both electrons and ions) has a significant effect, producing a force multiplier which transforms these sub-Eddington luminosities to being effectively super-Eddington.
Our model also shows the  disappearance of the wind in the hard state due to complete shadowing of the outer disc by the inner static corona. 
There is no requirement for a magnetic wind to explain the behaviour of the absorption features observed in this source. We will use Monte Carlo radiation transfer to explore the detailed spectra (as in \citealt{Tomaru2018}) in future work. 

The other black hole candidates which show absorption features from winds are likely also explained by thermal-radiative winds, 
especially those sources which have super-Eddington luminosities i.e. GRS 1915+105 and (probably) the anomalous wind seen in one observation of GRO J1655-40.

\section*{Acknowledgements}

Calculation of this work is performed of the XC30 and XC50 system 
at the the Center for Computational Astrophysics (CfCA), National Astronomical Observatory of Japan (NAOJ).
This work supported by JSPS KAKENHI Grant Number JP 19J13373 (RT), 
Society for the Promotion of Science Grant-in-Aid for Scientific Research (A) (17H01102 KO; 16H02170 TT), 
Scientific Research (C) (16K05309 KO; 18K03710 KO), 
and Scientific Research on Innovative Areas (18H04592 KO; 18H05463 TT).
This research is also supported by the Ministry of Education, Culture, Sports, Science and Technology of Japan as "Priority Issue on Post-K computer"(Elucidation of the Fundamental Laws and Evolution of the Universe) and JICFuS. 
RT acknowledges the support by JSPS Overseas Challenge Program for Young Resarchers.
CD acknowledges the Science and Technology Facilities Council (STFC) through grant ST/P000541/1, and visitor support from Kavli IPMU.

%%%%%%%%%%%%%%%%%%%%%%%%%%%%%%%%%%%%%%%%%%%%%%%%%%



\bibliographystyle{mnras}
\bibliography{library} % if your bibtex file is called example.bib

%\begin{thebibliography}{99}
%\end{thebibliography}

%%%%%%%%%%%%%%%%%%%%%%%%%%%%%%%%%%%%%%%%%%%%%%%%%%

%%%%%%%%%%%%%%%%% APPENICES %%%%%%%%%%%%%%%%%%%%%

\appendix 
\section{basic equation and numerical method}
\label{sec:code}

In this section, we include for completeness the full hydrodynamic equations. We solve these
in  spherical polar coordinates $(R,\phi, \theta)$.

The basic equations are the equation of continuity,
\begin{equation}
 \frac{\partial{\rho}}{\partial t} + \nabla \cdot (\rho \bm{v}) =0
 \label{eq:continuity}
\end{equation}
the equation of motion,
\begin{equation} %r 
 \frac{\partial (\rho v_R)}{\partial t} + \nabla \cdot (\rho v_R \bm{v}) =-\frac{\partial  p}{\partial R} +\rho( \frac{v^2_\mathrm{\theta}}{R}+ \frac{v^2_\mathrm{\phi}}{R}  + g_R+ f_\mathrm{\mathrm{rad}}(\xi, T) )
 \label{eq:vr}
\end{equation}
\begin{equation}%\theta
 \frac{\partial (\rho v_\mathrm{\theta})}{\partial t} + \nabla \cdot (\rho v_\mathrm{\theta} \bm{v}) =-\frac{1}{R}\frac{\partial  p}{\partial \theta} +\rho( -\frac{v_R v_\mathrm{\theta}}{R}+ \frac{v^2_\mathrm{\phi}}{R} \cot \theta  )
 \label{eq:vt}
\end{equation}
\begin{equation}
 \frac{\partial (\rho v_\mathrm{\phi})}{\partial t} + \nabla \cdot (\rho v_\mathrm{\phi} \bm{v}) = - \rho( \frac{v_\mathrm{\phi} v_R}{R}+ \frac{v_\mathrm{\phi} v_\mathrm{\theta}}{R} \cot \theta  )
 \label{eq:vp}
\end{equation}
and conservation of energy,
\begin{equation}
 \frac{\partial}{\partial t} \left[ \rho (\frac{1}{2} v^2+e) \right] + \nabla \cdot \left[ \rho \bm{v}(\frac{1}{2}v^2 +e + \frac{p}{\rho})\right]  =  \rho \bm{v} \cdot \bm{g}+\rho  \mathcal{L}(\xi, T)
 \label{eq:energy}
\end{equation}
where $\rho$ is the mass density, $\bm{v} = (v_R, v_\mathrm{\theta} , v_\mathrm{\phi}) $ is the velocity, 
$p$ is the gas pressure, $e$ is the internal energy per unit mass, and $\bm{g} = (g_R, 0)$ is 
the gravitational acceleration of the black hole. 
We assume an adiabatic equation of state $p/\rho = (\gamma - 1) e $ with ~$ \gamma = 5/3$. 
 
We set the computational domain from $R_\mathrm{in} = 0.01 R_\mathrm{out}\leq R \leq R_\mathrm{out}  $ , $0 \leq \theta \leq \pi/2$. We solve over $N_R=120$ grid points in $R$ and $N_\theta=240$ grid points in $\theta$.
The radial grid is set with geometric spacing as  
\begin{equation}
R_i = R_\mathrm{\mathrm{in}} (R_\mathrm{out}/R_\mathrm{in})^{i/N_R}, ~ \text{($0\leq i \leq N_R$)}
\end{equation}

We set the polar angular grid in two sections, one 
to follow the scale height of the disc to resolve the irradiated launch region
(Eq.~\ref{eq:scale_d}) and one to sample the wind behaviour over the 
rest of the domain. We define these using the 
angle from the mid-plain $\alpha_j =\pi/2 - \theta_\mathrm{N_\mathrm{\theta}-j}, (0\leq j \leq N_\mathrm{\theta})$  
\begin{equation}
\alpha_j  =
\begin{cases}
     \arctan \left[ f_d \left\{ (R_\mathrm{j}/R_\mathrm{\mathrm{out}})^{2/7}-(R_\mathrm{\mathrm{in}}/R_\mathrm{\mathrm{out}})^{2/7} \right\} \right]  ,&( {\scriptsize \text{$0\leq j \leq N_R$}}) \\
    \arcsin\left\{ \frac{1.0-\sin(\alpha_\mathrm{N_R})}{N_\mathrm{\theta}-N_R}(j-N_R)+\sin(\alpha_\mathrm{N_R}) \right\} , &({\scriptsize \text{$N_R < j \leq  N_\mathrm{\theta}$}})
\end{cases}
\label{eq:polar grid}
\end{equation}
where $\theta_0 = 0, \theta_\mathrm{N_\theta} = \pi/2 $. 
The disc surface is the top of Eq.\ref{eq:polar grid}. 
At each time step, the density of disc surface is updated via $n = L_x/(170 R^2)$, where $L_x$ is the filtered
continuum.  

We run {\sc cloudy} before the hydrodynamic code, changing the 
ionisation parameters and temperature to  generate tables of net heating/cooling 
rate $\mathcal{L}(\xi, T)$ and force multiplier $M(\xi, T)$ in order to input into the
hydrodynamic simulations. This assumes the optically thin limit, which is justified for the 
very highly ionised wind conditions which we are trying to reproduce here. 
The calculation grid of {\sc cloudy} is an $ 301 \times 121 $ logarithmically spaced grid $(\xi, T)$ 
in a domain $1.0\times 10^3\leq T \leq 1.0\times10^9 $ and $1.0\times 10^{-3} \leq \xi \leq 1.0\times 10^{12}$.

The hydrodynamic terms for an ideal fluid are solved using an approximate Reimann solver,
the HLL method \citep{Harten1983}.
We treat the radiation force as an explicit external force term using the force multiplier supplied 
via biliner interpolation from pre-calculated {\sc cloudy} table.
The numerical procedures are i) calculation of Eq.\ref{eq:continuity}--\ref{eq:vp} and Eq.\ref{eq:energy} 
except for the net heating/cooling term, 
ii) implicit update of temperature by zbrent root finding method \citep{Press1992} 
(see, also \citealt[A1.2]{Dyda2017}) using net heating/cooling rate via biliner interpolation 
from pre-calculated table by {\sc cloudy}.

The time step is determined using the Courant-Friendrichs-Levi condition. 
At each grid, we calculate 
\begin{equation}
    \delta t =0.3 \frac{\min(\delta R ,R\delta\theta)}{\sqrt{(v_R+c_s)^2+(v_\theta+c_s)^2}}
\end{equation}
where $\delta R , \delta \theta$ are the grid sizes, while  $c_s$ is the isothermal sound speed.
The minimum value of $\delta t$ in all grids is used as the time step.

\section{Density and temperature structure of all simulations}

\begin{figure*}
\includegraphics[width=0.8\hsize]{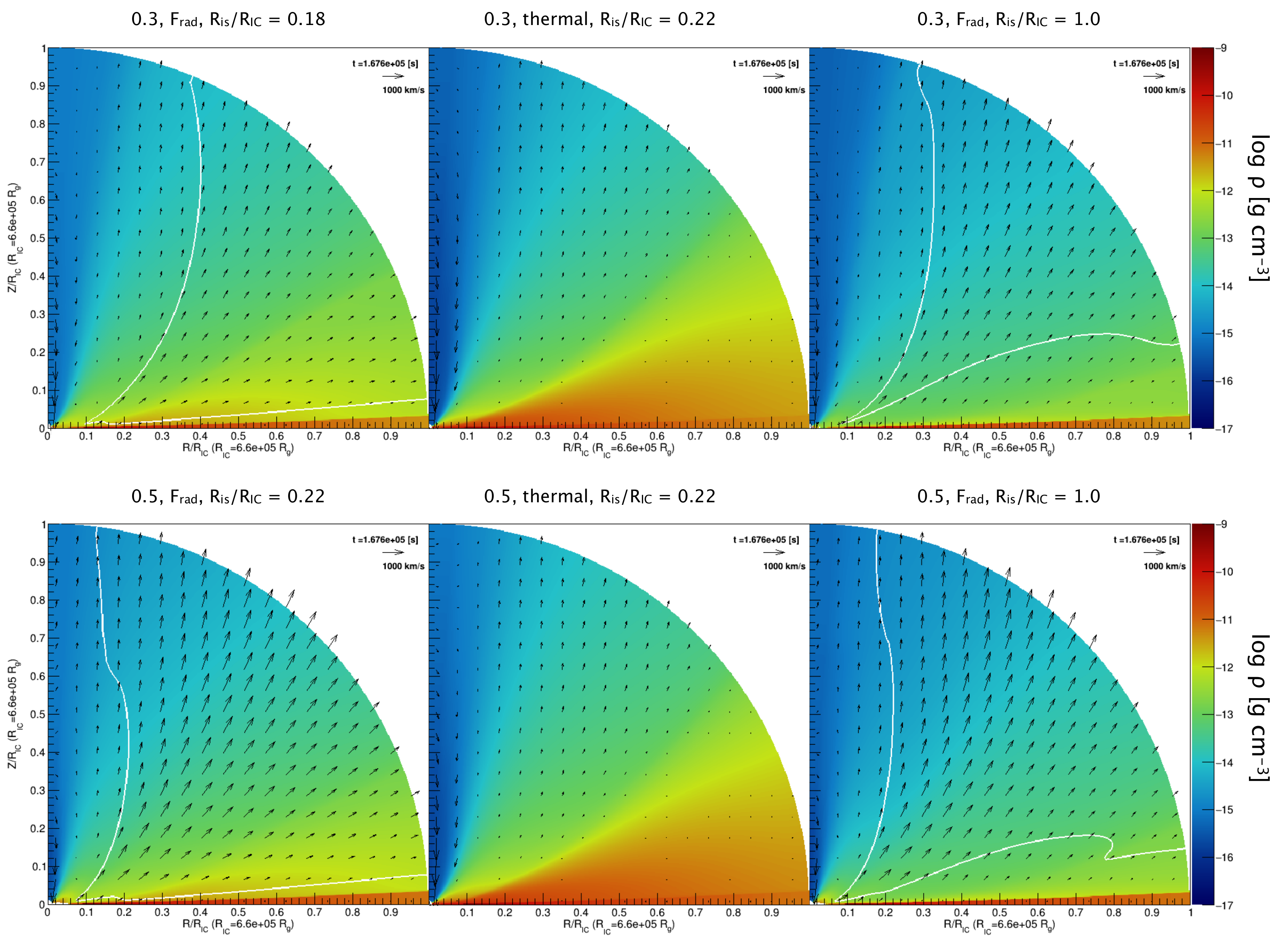}\\
\includegraphics[width=0.8\hsize]{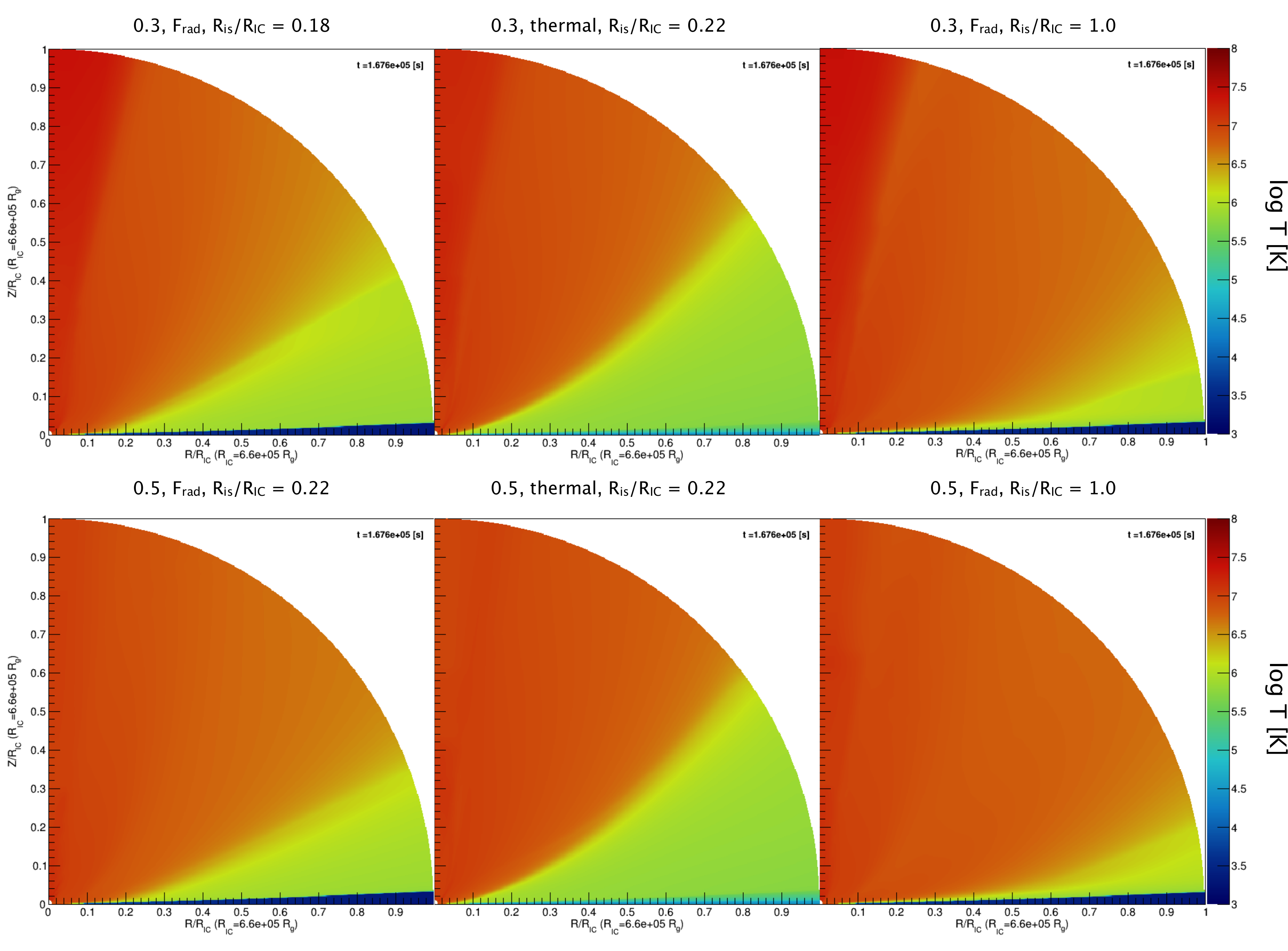}
\caption{The density (top) and temperature (bottom) structures of final state of each simulation. Solid white lines show the Mach 1 surface of the outflow ($v_{R}/c_{s} = v_{R}/\sqrt{\gamma k T/(\mu m_p)}=1$).}
\label{fig:temperature}
\end{figure*}

\begin{figure}
    \centering
    \includegraphics[width=0.9\hsize]{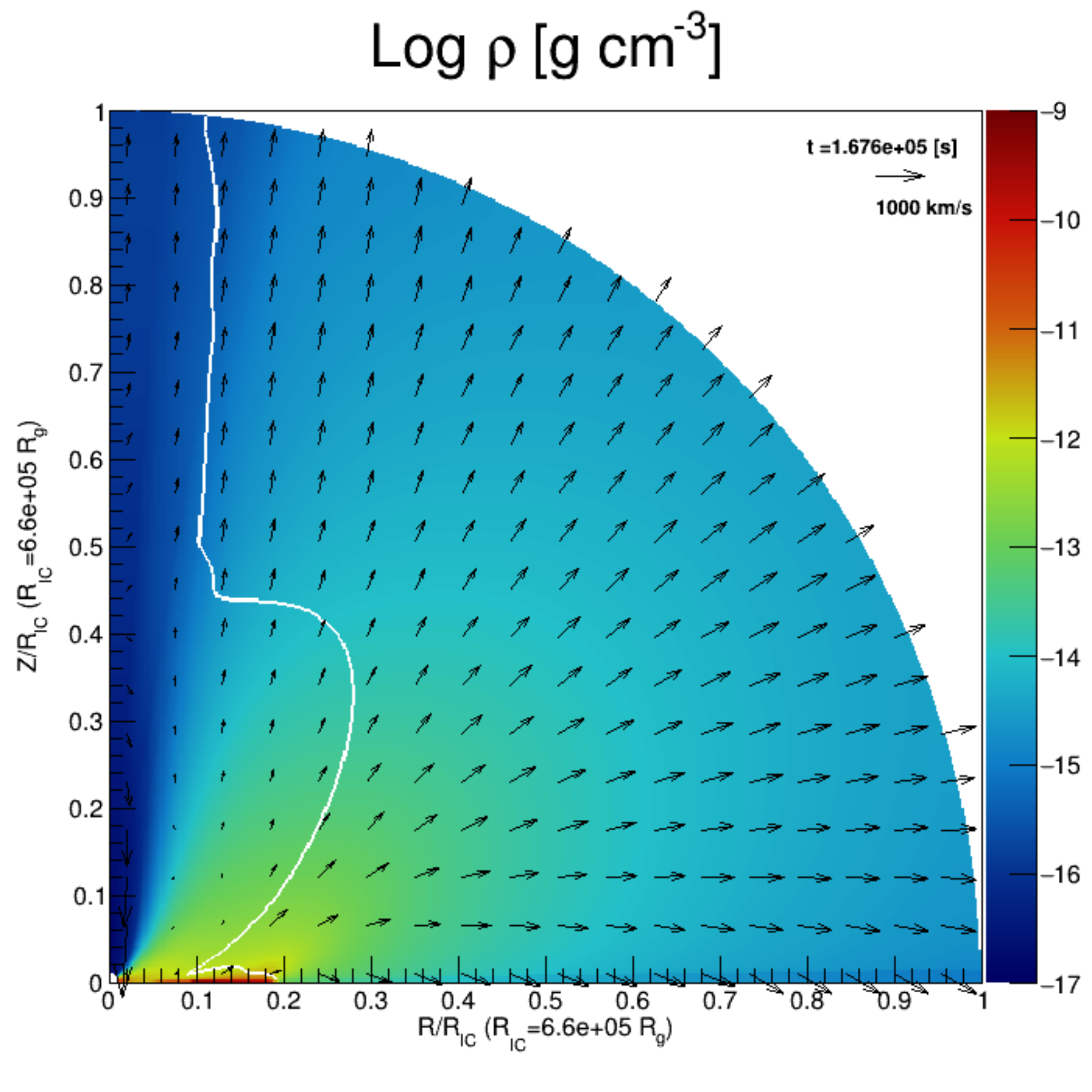}\\
    \includegraphics[width=0.9\hsize]{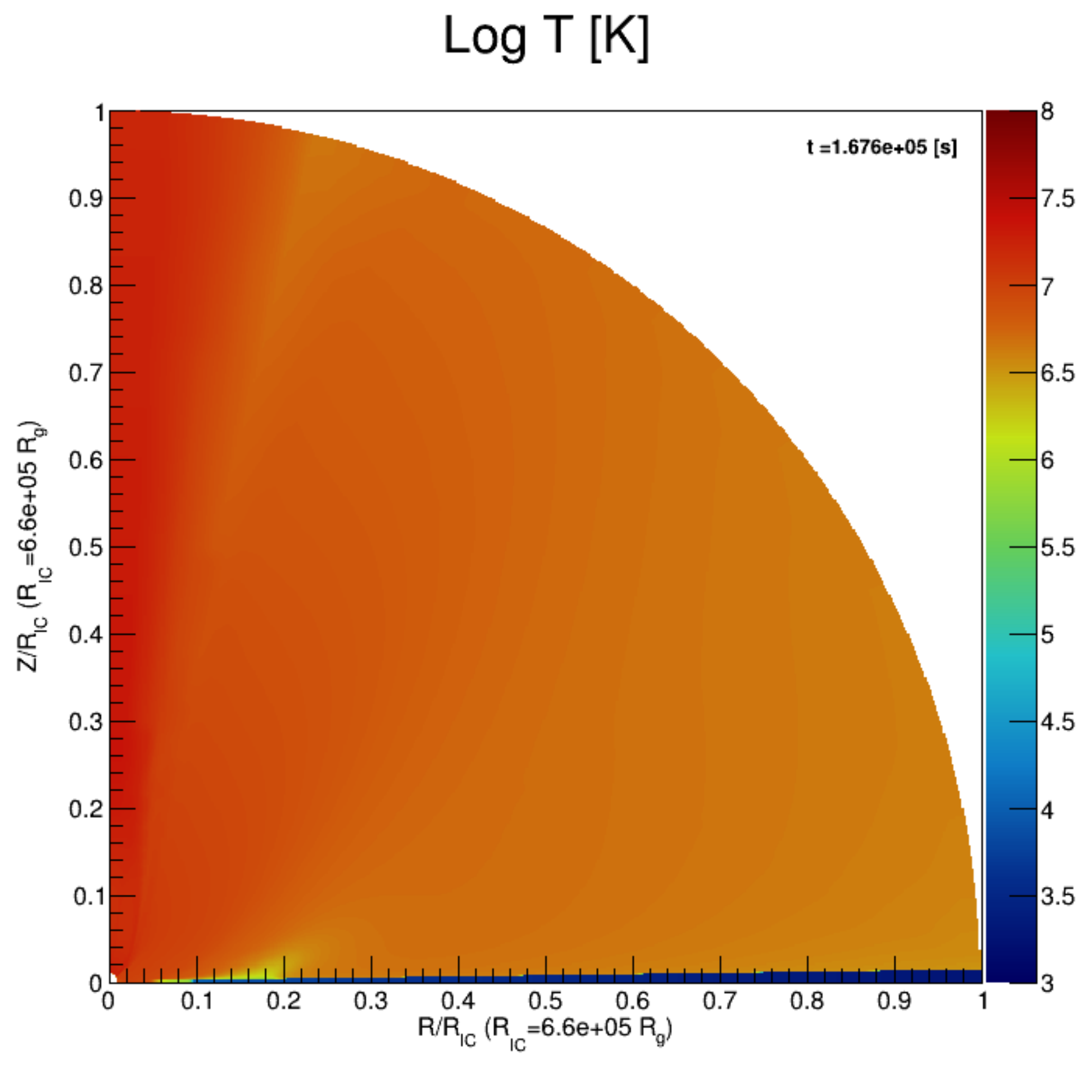}\\
    \caption{The density (top) and temperature (bottom) structure of small disc simulation.}
    \label{fig:small disc}
\end{figure}

\begin{figure}
    \centering
    \includegraphics[width=0.9\hsize]{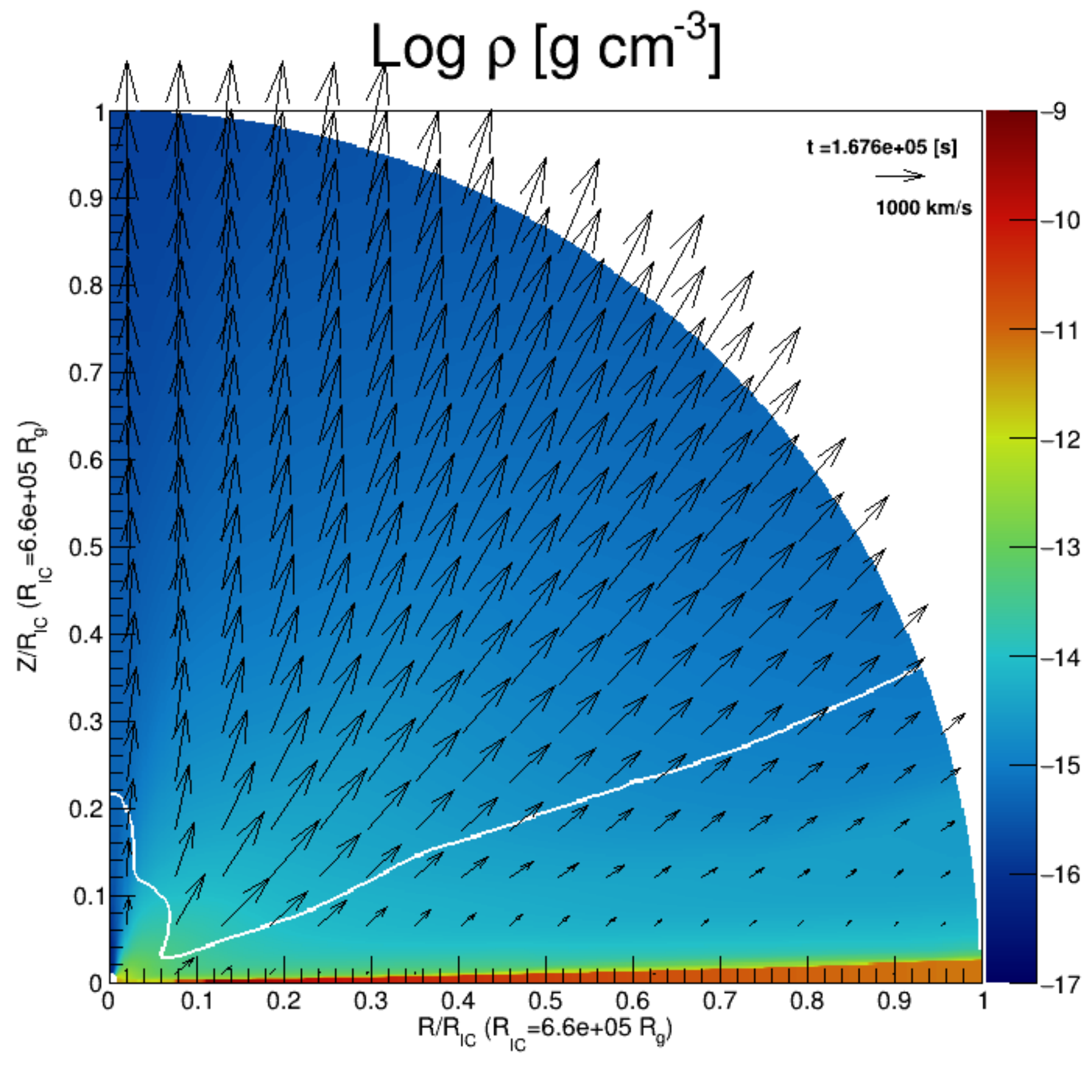}\\
    \includegraphics[width=0.9\hsize]{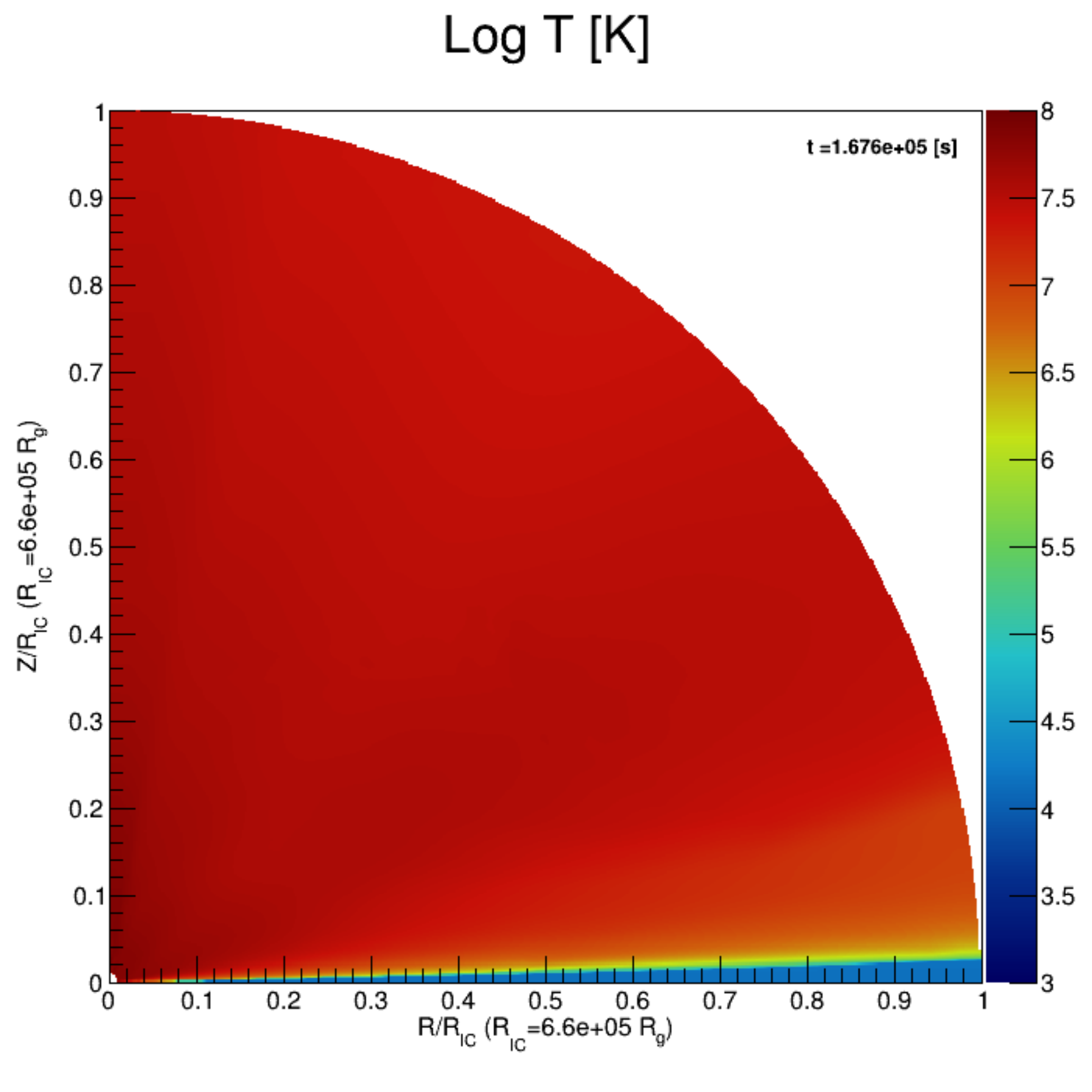}
    \caption{The density (top) and temperature (bottom) structure of hard-state }
    \label{fig:map hard-state}
\end{figure}

\begin{figure}
    \centering
    \includegraphics[width=0.9\hsize]{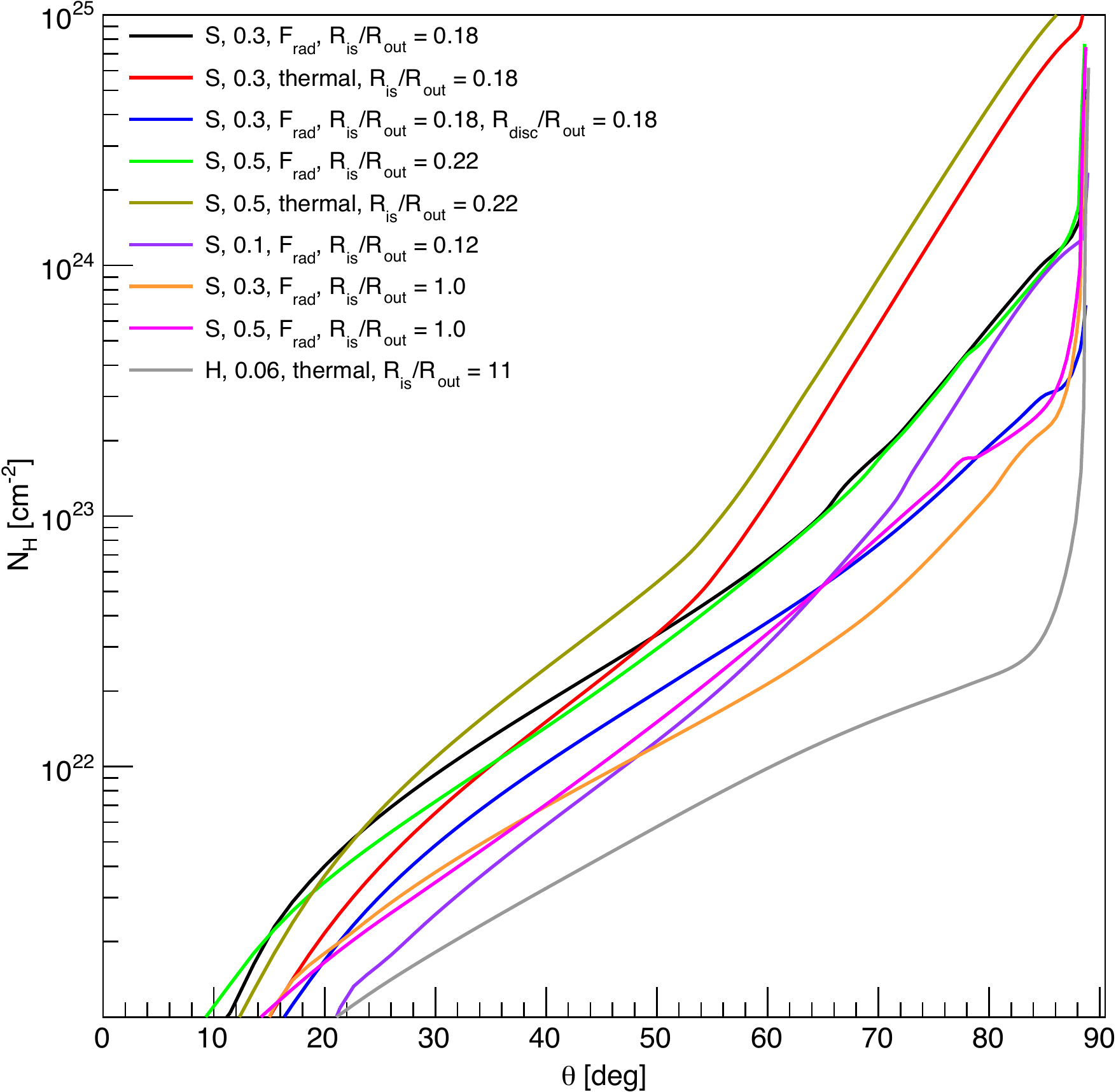}
    \caption{hydrogen column density of all simulations in Tab. \ref{tab:summary}}
    \label{fig:Nh}
\end{figure}

We show the density and temperature structure of simulations with $L/L_\text{LEdd} = 0.3, 0.5$ (Fig.\ref{fig:temperature}), that of hard state (Fig.\ref{fig:map hard-state}) and the all column densities in Tab. \ref{tab:summary}.

%At hydrodynamics times step, $\mathcal{L}(\xi, T), M(\xi, T)$ is update via
%biliner interpolation in $\log(\xi, T) $ space from these tables.
%This method to calculate energy equation is same as 
%\citet{Dyda2017,Higginbottom2015}
 
%The numerical procedure is devided into the follwing steps 

%This means that first 120 grid of $\alpha$ is used for disk, and another 120 grid is used for winds. 
%The each element of  physical quantities are calculated at grid middle points.

\if0
\section{Analysis of H~1743-322}
\begin{figure}
    \centering
    \includegraphics[width=0.9\hsize]{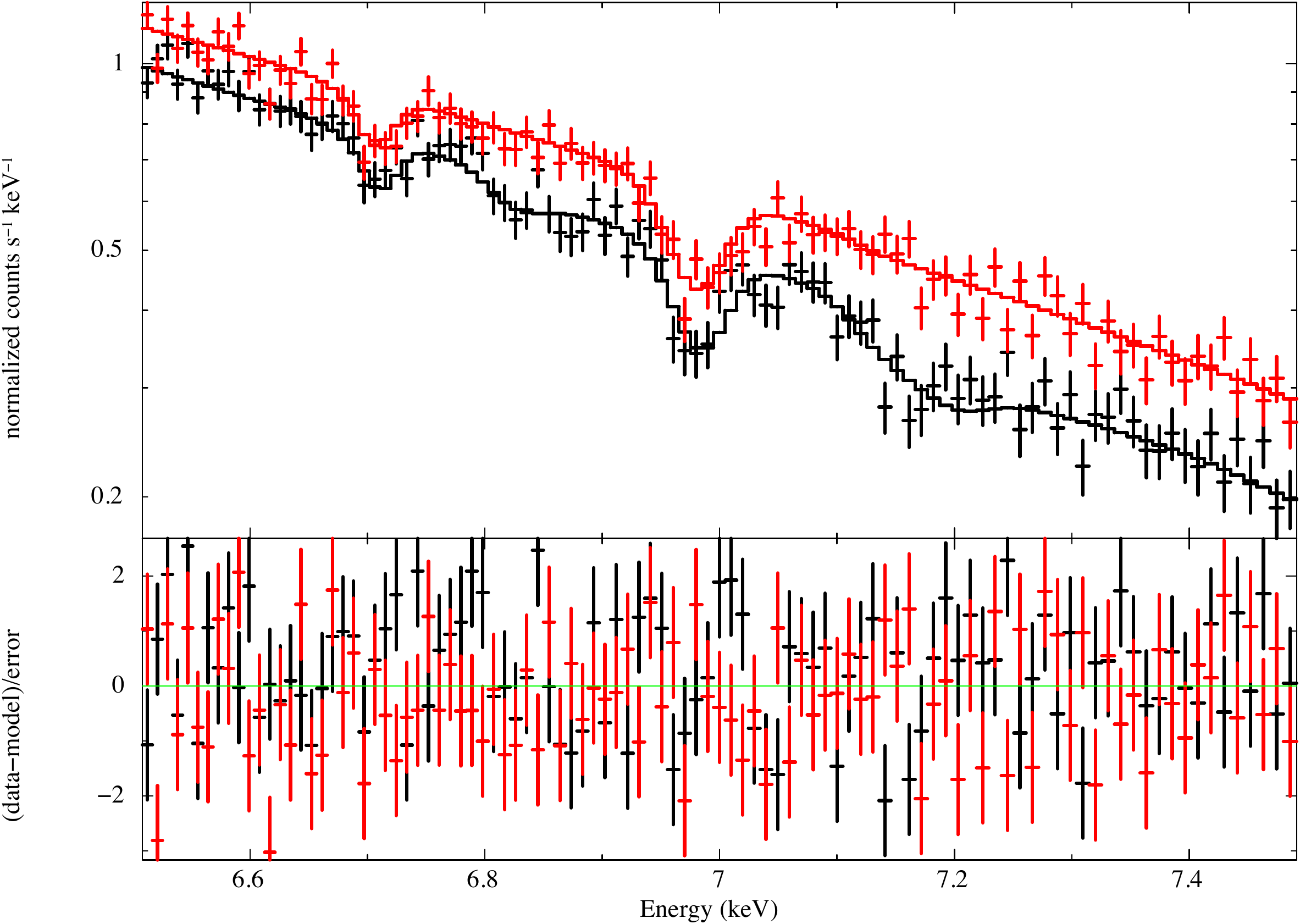}
    \caption{The absorption line of H~1743-322}
    \label{fig:obs}
\end{figure}
In this section, we breafly describe that re-analysis of the 

The last numbered section should briefly summarise what has been done, and describe
the final conclusions which the authors draw from their work.


If you want to present additional material which would interrupt the flow of the main paper,
it can be placed in an Appendix which appears after the list of references.
\fi 

%%%%%%%%%%%%%%%%%%%%%%%%%%%%%%%%%%%%%%%%%%%%%%%%%%

% Don't change these lines
\bsp	% typesetting comment
\label{lastpage}
\end{document}